\documentclass{article} %

\usepackage{iclr2024_conference,times}
\iclrfinalcopy

\usepackage{amsmath,amsfonts,bm}

\def\eqref#1{equation~\ref{#1}}

\def\1{\bm{1}}

\def\vg{{\bm{g}}}

\def\vl{{\bm{l}}}

\def\vx{{\bm{x}}}
\def\vy{{\bm{y}}}

\def\mX{{\bm{X}}}

\DeclareMathAlphabet{\mathsfit}{\encodingdefault}{\sfdefault}{m}{sl}
\SetMathAlphabet{\mathsfit}{bold}{\encodingdefault}{\sfdefault}{bx}{n}

\usepackage[utf8]{inputenc} %
\usepackage{graphicx}
\usepackage[T1]{fontenc}
\usepackage{hyperref}       %
\usepackage{url}            %
\usepackage{booktabs}       %
\usepackage{amsfonts}       %
\usepackage{nicefrac}       %
\usepackage{microtype}      %

\IfFileExists{headers/config/showoverfull.config}{
	\overfullrule=1cm
}{
}

\usepackage{marginnote}

\usepackage[backgroundcolor=none,linecolor=red,textsize=footnotesize]{todonotes}

\usepackage{etoolbox}

\newbool{includeappendix}
\setbool{includeappendix}{true} %
\IfFileExists{headers/config/noappendix.config}{
	\setbool{includeappendix}{false}
}{}

\newif\ifincludeappendixx
\ifbool{includeappendix}{
	\includeappendixxtrue
}{
	\includeappendixxfalse
}

\usepackage{xr} %
\usepackage{filecontents}

\ifbool{includeappendix}{}{
	\input{appendix-labels-loader}

	\externaldocument{appendix-labels}
}

\newcommand{\eg}{e.g., }
\newcommand{\ie}{i.e., }

\usepackage{acro} %

\usepackage{listings}

\usepackage{textcomp}

\usepackage{xcolor}

\usepackage[scaled=0.8]{beramono}

\definecolor{ckeyword}{HTML}{7F0055}
\definecolor{ccomment}{HTML}{3F7F5F}
\definecolor{cstring}{HTML}{2A0099}

\lstdefinestyle{numbers}{
	numbers=left,
	framexleftmargin=20pt,
	numberstyle=\tiny,
	firstnumber=auto,
	numbersep=1em,
	xleftmargin=2em
}

\lstdefinestyle{layout}{
	frame=none,
	captionpos=b,
}

\lstdefinestyle{comment-style}{
	morecomment=[l]//,
	morecomment=[s]{/*}{*/},
	commentstyle={\color{ccomment}\itshape},
}

\lstdefinestyle{string-style}{
	morestring=[b]",%
	morestring=[b]',%
	stringstyle={\color{cstring}},
	showstringspaces=false,%
}

\lstdefinestyle{keyword-style}{
	keywordstyle={\ttfamily\bfseries},
	morekeywords={
		function,
		constructor,
		int,
		bool,
		return,
		returns,
		uint
	},
	morekeywords = [2]{},
	keywordstyle = [2]{\text},
	sensitive=true,
}

\lstdefinestyle{input-encoding}{
	inputencoding=utf8,
	extendedchars=true,
	literate=
	{ℝ}{$\reals$}1%
	{→}{$\rightarrow$}1%
	{α}{$\alpha$}1%
	{β}{$\beta$}1%
	{λ}{$\lambda$}1%
	{θ}{$\theta$}1%
	{ϕ}{$\phi$}1%
}

\lstdefinestyle{escaping}{
	moredelim={**[is][\color{blue}]{\%}{\%}},
	escapechar=|,
	mathescape=true
}

\lstdefinestyle{default-style}{
	basicstyle=\fontencoding{T1}\ttfamily\footnotesize,
	style=numbers,
	style=layout,
	style=comment-style,
	style=string-style,
	style=keyword-style,
	style=input-encoding,
	style=escaping,
	tabsize=2,
	upquote=true
}

\lstdefinelanguage{BASIC}{
	language=C++,
	style=default-style
}[keywords,comments,strings]%

\usepackage{graphicx}
\usepackage{mathtools}%
\usepackage{enumitem}
\usepackage{bbm}    %
\usepackage{xspace}
\usepackage{textpos}	%
\usepackage[margin=4pt]{subcaption}	%
\usepackage{multirow}
\usepackage{algorithm}
\usepackage{algpseudocode}
\algtext*{EndFunction}

\usepackage[makeroom]{cancel} 
\usepackage{ stmaryrd }
\usepackage{wrapfig}
\usepackage{tabularx}
\usepackage{siunitx}

\newcommand{\tool}{SEER}
\newcommand{\toollong}{Data Stealing via \underline{Se}cret \underline{E}mbedding and \underline{R}econstruction}

\newcommand{\pd}[2]{\ensuremath{\left\lVert\frac{\partial{#1}}{\partial{#2}}\right\rVert}}

\newcommand{\param}{\ensuremath{{\bm \theta}}}

\newcommand{\yes}[0]{\mbox{\includegraphics[height=0.25cm]{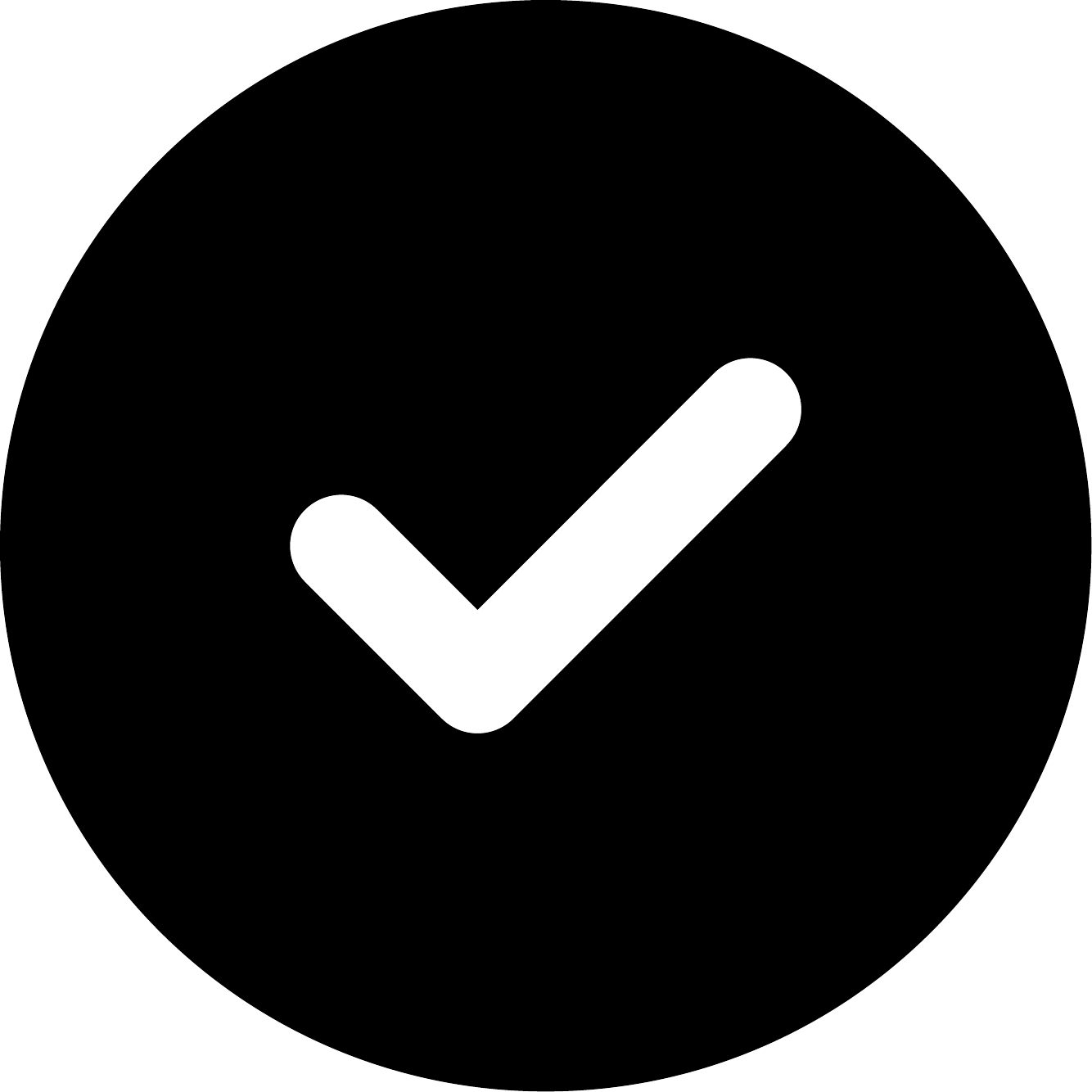}}\xspace}
\newcommand{\no}[0]{\mbox{\includegraphics[height=0.25cm]{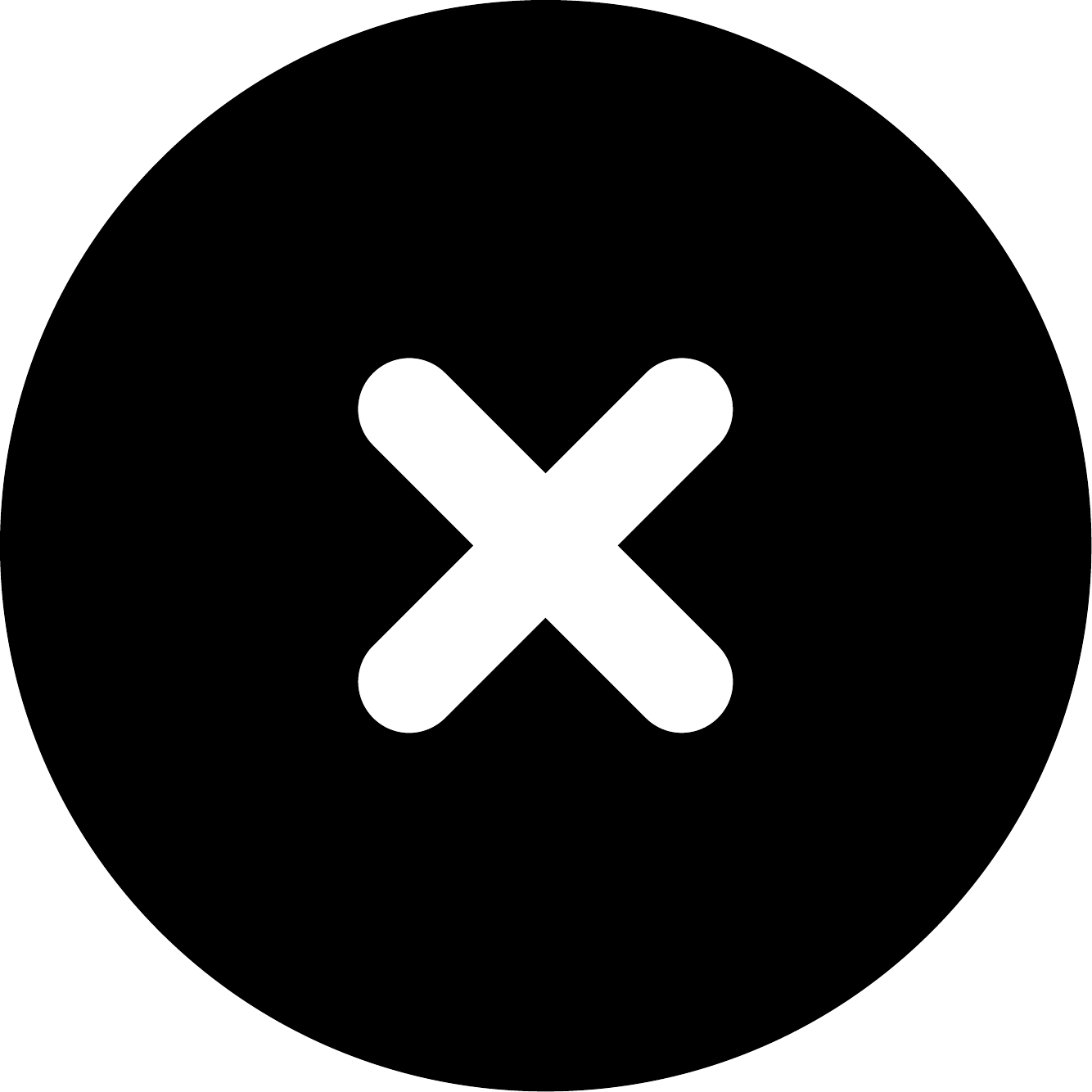}}\xspace}
\newcommand{\omg}[0]{\mbox{\includegraphics[height=0.25cm]{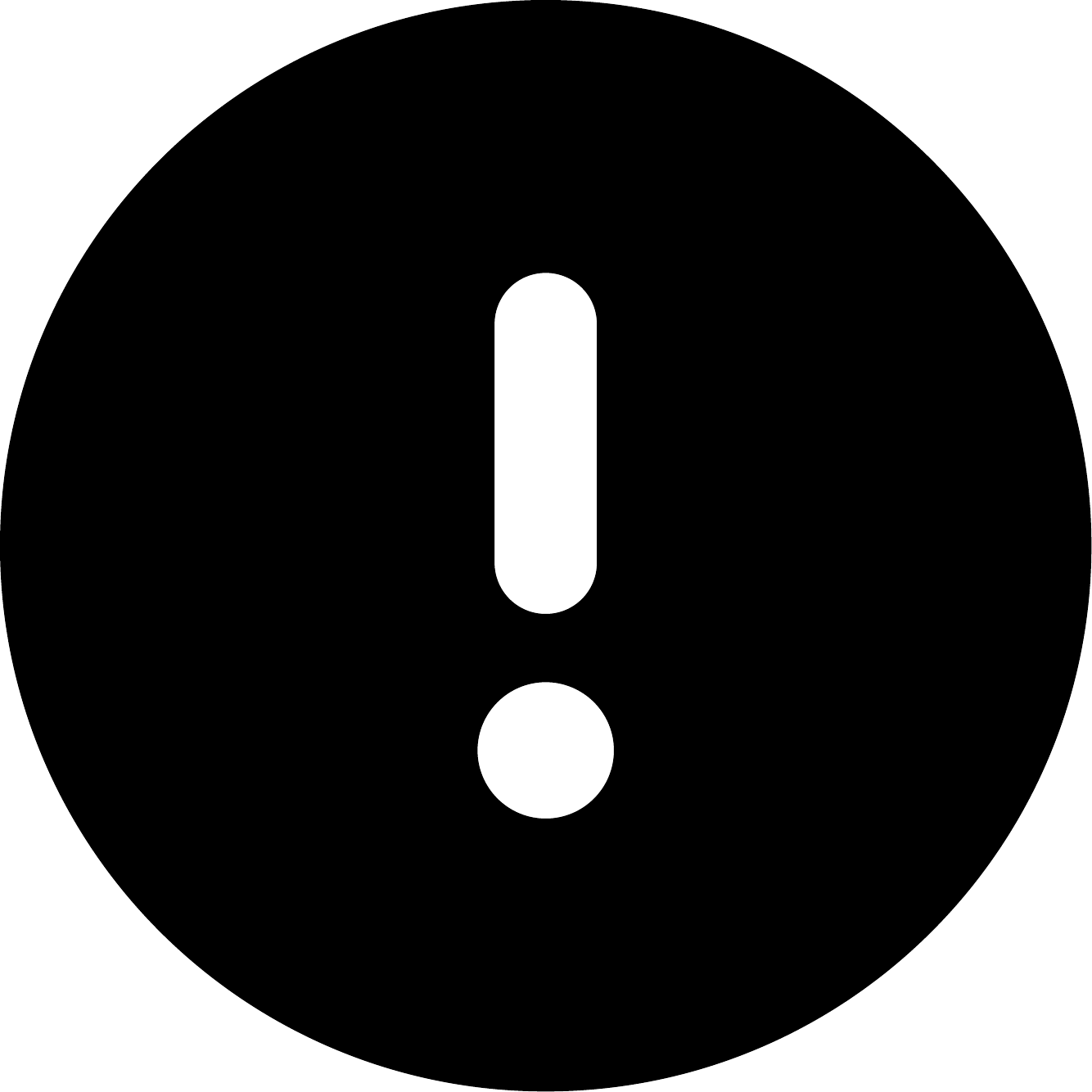}}\xspace}
\newcommand{\xblue}[0]{\mbox{\includegraphics[height=0.2cm]{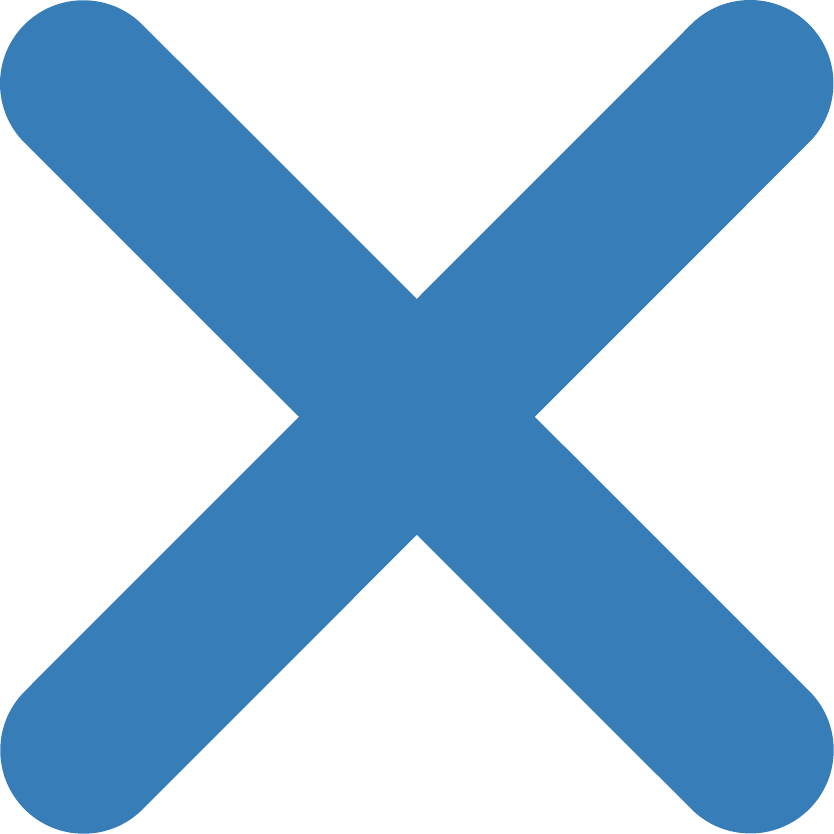}}\xspace}
\newcommand{\xred}[0]{\mbox{\includegraphics[height=0.2cm]{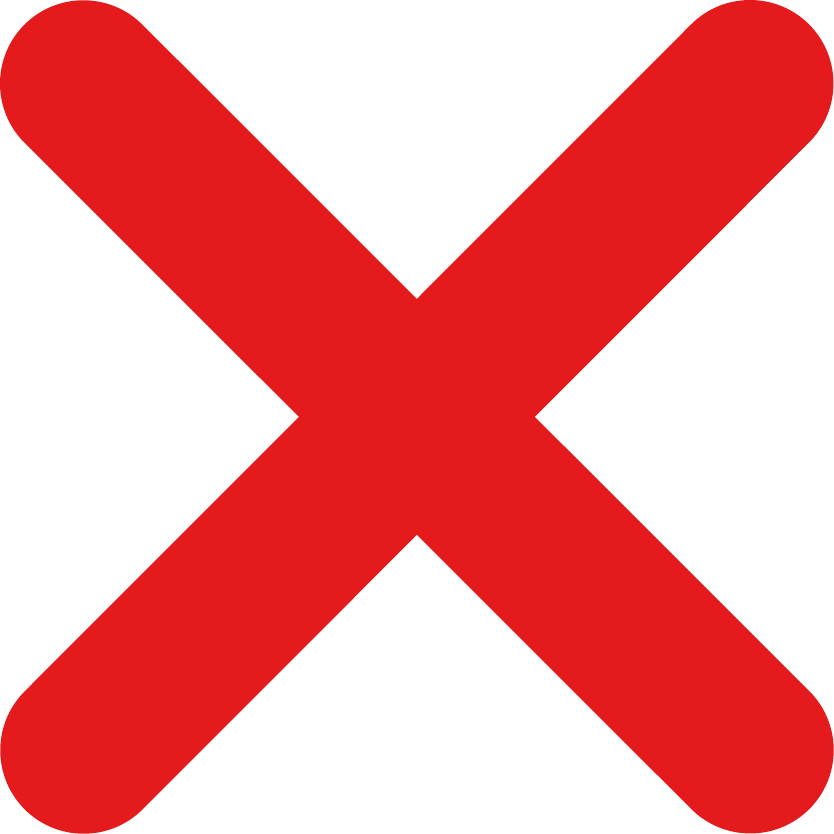}}\xspace}
\newcommand{\xgreen}[0]{\mbox{\includegraphics[height=0.2cm]{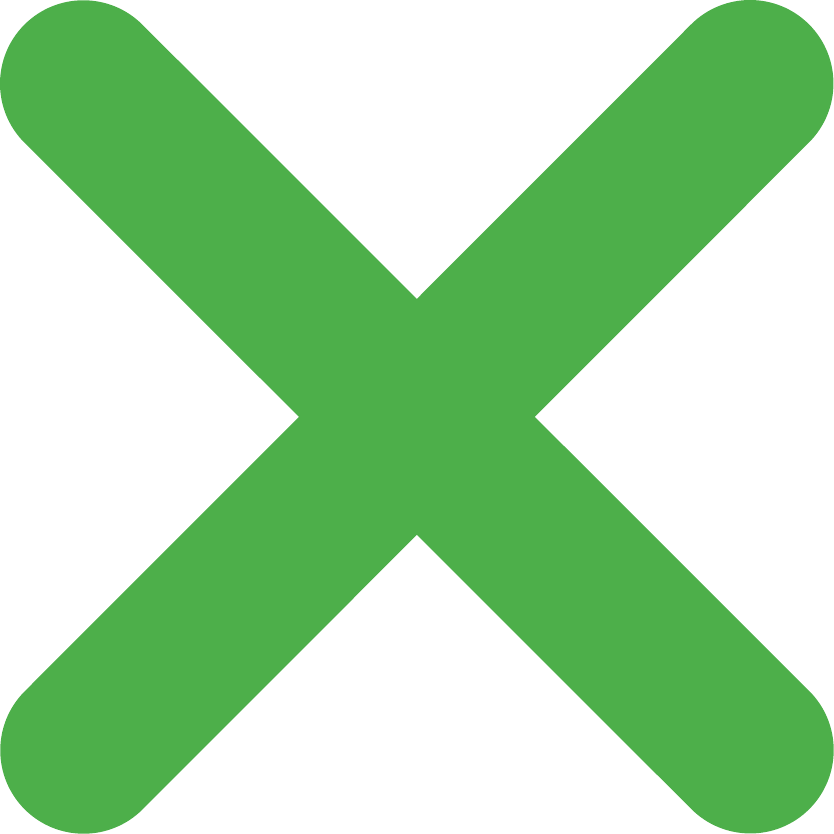}}\xspace}

\setcitestyle{square}

\usepackage[capitalize,noabbrev]{cleveref}

\hypersetup{
    colorlinks,
    linkcolor={black},
}

\crefformat{section}{\S#2#1#3}

\crefrangeformat{section}{\S#3#1#4\crefrangeconjunction\S#5#2#6}

\crefmultiformat{section}{\S#2#1#3}{\crefpairconjunction\S#2#1#3}{\crefmiddleconjunction\S#2#1#3}{\creflastconjunction\S#2#1#3}

\newcommand{\crefrangeconjunction}{--}

\crefname{listing}{Lst.}{listings}
\crefname{line}{Line}{Lines}
\crefname{appendix}{App.}{App.}

\newcommand{\appref}[1]{%
	\ifbool{includeappendix}{\cref{#1}}{the appendix}%
}
\newcommand{\Appref}[1]{%
	\ifbool{includeappendix}{\cref{#1}}{The appendix}%
}

\definecolor{refcolor}{RGB}{23, 120, 108} 
\definecolor{citcolor}{RGB}{23, 120, 18}

\crefformat{figure}{Fig.~#2{\color{refcolor}#1}#3}
\Crefformat{figure}{Fig.~#2{\color{refcolor}#1}#3}
\crefformat{algorithm}{Algorithm~#2{\color{refcolor}#1}#3}
\crefformat{line}{Line~#2{\color{refcolor}#1}#3}
\Crefformat{line}{Line~#2{\color{refcolor}#1}#3}
\crefrangeformat{line}{Lines~(#3{\color{refcolor}#1}#4--#5{\color{refcolor}#2}#6)}
\crefformat{appendix}{App.~#2{\color{refcolor}#1}#3}
\Crefformat{appendix}{App.~#2{\color{refcolor}#1}#3}
\crefformat{equation}{Eq.~#2{\color{refcolor}#1}#3} %
\Crefformat{equation}{Eq.~#2{\color{refcolor}#1}#3} %
\crefformat{section}{Sec.~#2{\color{refcolor}#1}#3}
\Crefformat{section}{Sec.~#2{\color{refcolor}#1}#3}
\crefformat{table}{Table~#2{\color{refcolor}#1}#3}
\Crefformat{table}{Table~#2{\color{refcolor}#1}#3}
\crefrangeformat{table}{Tables~#3{\color{refcolor}#1}#4 to~#5{\color{refcolor}#2}#6}
\Crefrangeformat{table}{Tables~#3{\color{refcolor}#1}#4 to~#5{\color{refcolor}#2}#6}

\hypersetup{citecolor=citcolor}

\title{Hiding in Plain Sight: Disguising Data \\ Stealing Attacks in Federated Learning}

\author{Kostadin Garov$^{1}$\hspace{2.8em}Dimitar I. Dimitrov$^{1,2}$\hspace{2.8em}Nikola Jovanovi\'c$^{2}$\hspace{2.8em}Martin Vechev$^{2}$\\
	\hfil$^{1}$ INSAIT, Sofia University "St. Kliment Ohridski"\hspace{3em}$^{2}$ ETH Zurich\hfil\\
	\hfil\texttt{\{kostadin.garov, dimitar.iliev.dimitrov\}@insait.ai} $^{1}$\hfil\\
	\hfil\texttt{\{nikola.jovanovic, martin.vechev\}@inf.ethz.ch} $^{2}$\hfil\\
}

\begin{document}
\maketitle

\vspace{-1em}
\begin{abstract}
Malicious server (MS) attacks have enabled the scaling of data stealing in federated learning to large batch sizes and secure aggregation, settings previously considered private. However, many concerns regarding the client-side detectability of MS attacks were raised, questioning their practicality. In this work, for the first time, we thoroughly study client-side detectability. We first demonstrate that all prior MS attacks are detectable by principled checks, and formulate a necessary set of requirements that a practical MS attack must satisfy. Next, we propose \tool{}, a novel attack framework that satisfies these requirements. The key insight of \tool{} is the use of a secret decoder, jointly trained with the shared model. We show that \tool{} can steal user data from gradients of realistic networks, even for large batch sizes of up to 512 and under secure aggregation. Our work is a promising step towards assessing the true vulnerability of federated learning in real-world settings.
\end{abstract}
\vspace{-1em}

\section{Introduction} \label{sec:intro}

Federated learning~(FL, \citet{fedsgd}) was proposed as a way to train machine learning models while preserving client data privacy. Recently, FL has seen a dramatic increase in real-world deployment~\citep{gboard,siri,webank}.
In FL, a \emph{server} trains a \emph{shared model} by applying aggregated gradient updates, received from numerous \emph{clients}.

\paragraph{Gradient leakage attacks}
A long line of work~\citep{dlg,geiping,rgap,aaai,nvidia} has shown that even passive servers can reconstruct client data from gradients, breaking the key privacy promise of FL.
However, these attacks are only applicable to naive FL deployments~\citep{arora}---in real-life settings with no unrealistic assumptions, they are limited to small batch sizes with no secure aggregation~\citep{secagg}.
In response, recent work has argued that the honest-but-curious threat model underestimates the risks of FL, as real-world servers can be malicious or compromised.  
This has led to \emph{malicious server (MS)} attacks, which have demonstrated promising results by lifting honest attacks to large batch sizes. 

Most prior MS attacks rely on one of two key underlying principles.
One attack class~\citep{cah,rtf,mandrake,zhang23} uses malicious model modifications to encourage sparsity in dense layer gradients, enabling the application of analytical honest attacks---we refer to these as \emph{boosted analytical} attacks.
Other attacks utilize \emph{example disaggregation}~\citep{pasquini,fishing}, reducing the effective batch size in the gradient space by restricting gradient flow, which permits the use of optimization-based honest attacks.

\paragraph{Client-side detectability}
Nearly all prior work in the field~\citep{geiping,cah,rtf,pasquini,fishing,decepticons,panning,mandrake} raised the issue of \emph{client-side detectability} of MS attacks, \ie an FL client may be able to detect malicious server activity, and decide to opt out of the current or future rounds.
Despite such concerns, no attempts were made to study, quantify, or reduce the detectability of MS attacks.  

\paragraph{This work: detecting and disguising malicious server attacks}
We thoroughly study the question of client-side detectability of MS attacks.
We demonstrate that while boosted analytical and example disaggregation attacks pose a real threat as zero-day exploits, now that their key principles are known, \emph{all} current (and future) attacks from these two classes are client-side detectable in a principled manner, bringing into question their practicality.
Notably, we demonstrate the detectability of (the more promising) example disaggregation attacks by introducing D-SNR, a novel vulnerability metric.

We observe that such limitations of prior MS attacks arise from their fundamental reliance on the honest attacks they lift.
Namely, boosted analytical attacks always require handcrafted modifications which are \emph{weight space detectable}, and example disaggregation attacks rely on the success of disaggregation, which is equally evident to any party observing the gradients, \ie it is \emph{gradient space detectable}.
This illustrates the need for fundamentally different attack approaches.

As a step in that direction, we propose a novel attack framework \tool{}, which recovers data from batch sizes up to 512, yet is by design harder to detect than prior attacks.
Our key insights are that (i) gradient space detection can be evaded using a \emph{secret decoder}, disaggregating the data in a space unknown to clients, and (ii) jointly optimizing the decoder and the shared model with SGD on auxiliary data \emph{avoids handcrafted modifications} and allows for effective reconstruction.
Importantly, \tool{} does not lift any prior honest attack and does not require restrictive assumptions such as architecture tweaking, side-channel information, or knowledge of batch normalization data or labels.

\paragraph{Key contributions}
Our main contributions are:
\begin{itemize}
    \item We demonstrate that both boosted analytical and example disaggregation MS attacks are detectable using principled checks---for the latter, we introduce D-SNR, a novel gradient space metric of data vulnerability that can protect clients from unintended leakage. We formulate a necessary set of requirements for realistic MS attacks and make the case that detection should become a key concern when designing future attacks (\cref{sec:detection}).
    \item We propose \tool{}, a novel attack framework which satisfies all requirements based on malicious training of the shared model with a secret server-side decoder. \tool{} is harder to detect by design as it does not rely on honest attacks, avoiding previous pitfalls (\cref{sec:attack}). We provide an implementation of SEER at \href{https://github.com/insait-institute/SEER}{https://github.com/insait-institute/SEER}.
    \item We present an extensive experimental evaluation of \tool{} on several datasets and realistic network architectures, demonstrating that it is able to recover private client data from batches as large as 512, even under the presence of secure aggregation (\cref{sec:experiments}).
\end{itemize} 
\section{Related Work} \label{sec:related}
In this section, we discuss prior work on gradient leakage attacks in federated learning.
\paragraph{Honest server attacks} 
\emph{Optimization-based attacks}~\citep{dlg, iDLG, geiping, aaai, lti, nvidia} optimize a dummy batch to match the user gradient.
\emph{Analytical attacks}~\citep{ analyticPhong, cpa} recover inputs of linear layers in closed form, but are limited to batch size $B=1$ and do not support convolutional networks.
\emph{Recursive attacks}~\citep{ rgap} extend analytical attacks to convolutional networks but are limited to $B \leq 5$.
Several works thoroughly study all three attack classes~\citep{ study,bayesian,cafe,arora}.
Crucially,~\citet{arora} show that in realistic settings, where clients do not provide batchnorm statistics and labels, all honest attacks are limited to $B < 32$ for low-res data, and fail even for $B=1$ on high-res data.
This implies that large $B$ and secure aggregation~\cite{secagg} are effective protections against honest attacks.
 
\paragraph{Malicious server (MS) attacks} 
We focus on broadly applicable boosted analytical~\citep{cah,rtf,mandrake,zhang23} and example disaggregation attacks~\citep{fishing,pasquini}, discussed in~\cref{sec:detection}. 
Here, we reflect on other MS attacks that study more specific or orthogonal settings.
Several studies~\citep{pasquini,mandrake} require the ability to send a different update to each user, which was shown easy to mitigate with reverse aggregation~\citep{pasquini}.
\citet{lam} focuses on the rare setting with participation side-channel data.
While we target image reconstruction, some works consider other modalities, such as text~\citep{ LAMP,princeton,decepticons,panning} or tabular data~\citep{ oldtabular, tableak}.
Further, while we focus on the threat of data reconstruction,~\citet{ pasquini} studies weaker privacy notions such as membership~\citep{ mia} or property inference~\citep{ propertyInf}.
Finally, sybil-based attacks are a notably stronger threat model orthogonal to our work~\citep{sybils,boenisch23}. We further detail our exact threat model in \cref{app:threat_model}.
\section{Detectability of Existing Malicious Server Attacks} \label{sec:detection}
Most malicious server (MS) attacks rely on one of two strategies based on which we group them into two classes---\emph{boosted analytical} and \emph{example disaggregation}.
We now discuss client-side detectability of these classes and show that both are detectable with principled checks. We identify the root cause of detectability and formulate necessary requirements that future attacks must satisfy to be practical.

\paragraph{Boosted analytical attacks} %
The works of \citet{cah}, \citet{rtf}, \citet{mandrake}, and \citet{zhang23} use model modifications to induce different variants of {sparsity in dense layer gradients}, enabling the application of honest analytical attacks to batch sizes beyond one. 
Applying such attacks to the realistic case of convolutional networks requires highly unusual \emph{architectural modifications}, \ie placing a large dense layer in front, which makes the attack obvious.
The only alternative way to apply these attacks is to set all convolutions to identity, such that the inputs are transmitted unchanged to the dense layer.
As this is a pathological case that never occurs naturally and requires handcrafted changes to almost all parameters (\eg 98\% of weights in ResNet18), this approach is easily detectable by inspecting model weights (\eg by searching for convolutional filters with a single nonzero entry, see~\cref{sec:app_detection}).
More importantly, high levels of transmission are, in fact, impossible in realistic networks due to pooling and strides~\citep{rtf}, and further attempts to conceal the changes (\eg by adding weight noise) would additionally worsen the results.

\paragraph{Example disaggregation attacks} 
While the detectability of boosted analytical attacks was recognized in prior work~\citep{geiping,cah,fishing}, example disaggregation attacks~\citep{fishing,pasquini} are considered more promising.
These attacks use model modifications to restrict the gradient flow for all but one example, causing the aggregated gradient of a batch to be equal to the gradient of a single example. 
This undoes the protection of aggregation and allows the attacker to apply honest optimization-based attacks to reconstruct that example. 
While most instantiations of example disaggregation attacks rely on unusual handcrafted parameter changes, which are detectable in the \emph{weight space} (as for boosted analytical attacks), it may be possible to design variants that better disguise the gradient flow restriction.
For this reason, we focus on a more fundamental limitation of all (current and future) example disaggregation attacks and demonstrate it makes them easily detectable in the \emph{gradient space}.
Moreover, such detection is possible without running costly optimization-based attacks by using a simple principled metric.

We now propose one such metric, the \emph{disaggregation signal-to-noise ratio (D-SNR)}.
Assuming the use of the standard cross-entropy loss $\mathcal{L}(x, y)$, a shared model with parameters ${\bm \theta}$, and a batch of data $D = \{(x_1, y_1), \ldots, (x_B, y_B)\}$, we define D-SNR as follows:
\begin{equation} \label{eq:dsnr}
    {D\hbox{-}SNR}({\bm \theta}, D) = \max_{W \in {\bm \theta}_{lw}} \frac{\max_{i \in \{1,\ldots,b\}} \pd{\mathcal{L}(x_i, y_i)}{W}}{\sum_{i=1}^b \pd{\mathcal{L}(x_i, y_i)}{W} - \max_{i \in \{1,\ldots,b\}}\pd{\mathcal{L}(x_i, y_i)}{W}}
\end{equation}
where ${\bm \theta}_{lw}$ denotes the set of weights of all linear layers (dense and convolutional; 98\% of ResNet18).
Intuitively, D-SNR searches for layers where the batch gradient (the average of example gradients) is dominated by the gradient of a single example, suggesting disaggregation.
We conservatively use $\max$ to avoid false negatives and account for attempts at partial disaggregation, \ie if there is \emph{any} layer that disaggregates a single example, D-SNR will be large, and the client may decide that their batch is vulnerable and skip the current training round.
While we focus on the case of disaggregating a single example, our approach can be easily generalized to any number of examples.

\begin{figure*}[t]  
	\centering
	\vspace{-2em}
	\includegraphics[width= 0.91 \textwidth]{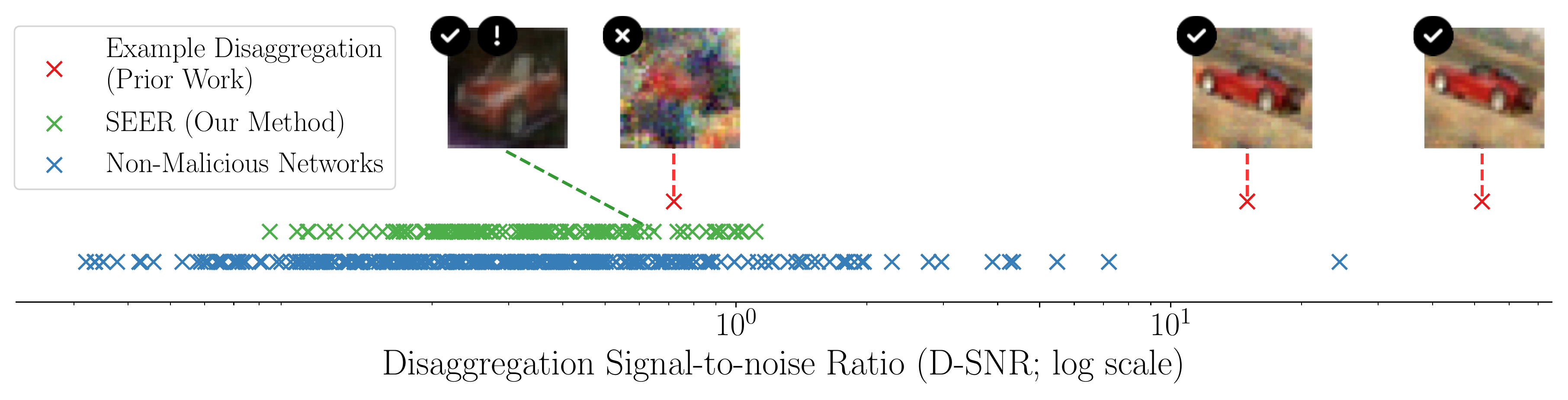}
	\vspace{-0.7em}
	\caption{D-SNR (\cref{sec:detection}) of real (model, data batch) pairs. High values indicate vulnerability to data leakage, which can manifest even in non-malicious models (\xblue). Example disaggregation attacks (\xred) are easily detectable as they can successfully reconstruct data (\protect\yes) only when DSNR is unusually high (note the log scale), and fail otherwise (\protect\no). Our method, \tool{} (\xgreen,~\cref{sec:attack}), successfully reconstructs an example even when D-SNR is low (\protect\yes \protect\omg), and is thus hard to detect in the original gradient space.
	\vspace{-1.2em}
	}
	\label{fig:dsnr}
\end{figure*}

We use D-SNR to experimentally study the detectability of example disaggregation attacks in realistic settings (see~\cref{sec:app_detection} for experimental details). 
As D-SNR is always $\infty$ for attacks proposed by~\citet{fishing}, we modify them in an attempt to smoothly control the strength of the gradient flow restriction.
Our key observation, presented in~\cref{fig:dsnr} (red \xred), is that in all cases where the attack is successful, D-SNR is unusually large, making the attack easily detectable. Reducing the strength of the gradient flow restriction further 
causes a sharp drop in D-SNR, entering the range of most non-malicious networks (blue \xblue), \ie the attack is undetectable. However, in all such cases, the attack fails, as the aggregation protects the examples.
In rare cases (\eg when overfitting), even natural networks can produce high D-SNR and be flagged. 
This behavior \emph{is desirable}, as such networks indeed disaggregate a single example, and (unintentionally) expose sensitive user data.
Thus, metrics such as D-SNR should be interpreted as detecting \emph{vulnerability}, and not necessarily \emph{maliciousness}. 

\paragraph{Requirements for future attacks}
Our results show that all prior MS attacks are client-side detectable using generic checks.
We argue that this is caused by fundamental problems in the design principles of the two attack classes and can not be remedied by further refinements. Any attempt to lift an honest analytical attack will inherit the limitation of being inapplicable to convolutions and will {require} architectural changes or handcrafted modifications detectable in the weight space.
Lifting optimization-based attacks always requires example disaggregation, which is gradient space detectable. 
More broadly, as all information needed to execute these attacks is in the user gradients, the server has no informational advantage and no principled way to conceal the malicious intent.

This suggests that new attack principles are required to better exploit the potential of the MS threat model.
To help guide the search, we now state the necessary requirements for future MS attacks guided by our results above and observations from prior work~\citep{fishing,arora}.
We argue that realistic MS data stealing attacks for image classification should: (i) target realistic deep convolutional networks with large batch sizes and/or secure aggregation; (ii) only utilize the attack vector of weight modifications, with no protocol changes (\eg non-standard architectures, asymmetric client treatment) and no sybil capabilities; (iii) not assume unrealistic side information, such as batch normalization statistics or label information~\citep{arora}; and (iv) explicitly consider the aspects of weight and gradient space detection (\eg avoid obvious handcrafted modifications).     
\section{\tool{}: \toollong{}} \label{sec:attack}  

In this section, we propose \tool, a novel attack framework that steals data from large batches while satisfying the requirements in~\cref{sec:detection}.
\tool{} avoids both pitfalls of prior MS attacks that caused them to be detectable.
Namely, \tool{} does not lift any honest attack and evades gradient space detection by disaggregating the data in a \emph{hidden space} defined by a server-side \emph{secret decoder}.
As a result, \tool{}-trained networks (green \xgreen) have D-SNR values indistinguishable from those of natural networks (\cref{fig:dsnr}). Further, \tool{} does not use handcrafted modifications, and instead trains the shared model and the secret decoder jointly with SGD, evading weight space detection.

\begin{figure}[t]
	\centering
	\vspace{-2em}
	\includegraphics[width=\textwidth]{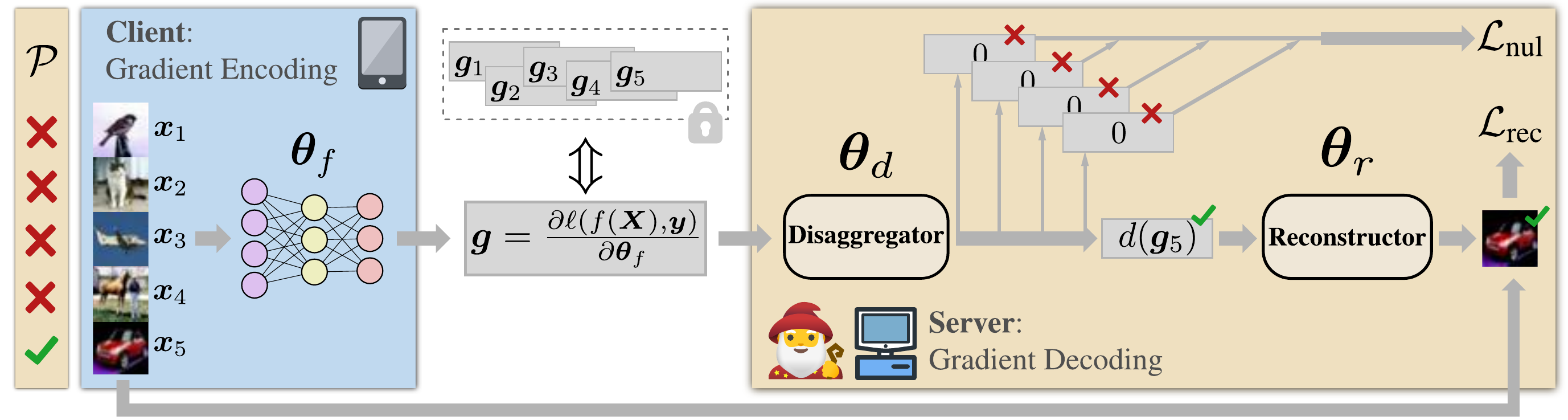}
	\caption{Overview of \tool{}. A client propagates a batch ${\mX}$ (of which one image satisfies the property $\mathcal{P}$ only known to the server) through the shared network $f$ with malicious weights $\param_f$, and returns the aggregated gradient $\vg$, hoping that the aggregation protects individual images. The server steals the image satisfying $\mathcal{P}$ by applying a secret disaggregator $d$ to remove the impact of other images in a hidden space, followed by a secret reconstructor $r$. \tool{} is trained by jointly optimizing $\param_f$, $\param_d$, and $\param_r$ to minimize a weighted sum of $\mathcal{L}_{\text{nul}}$ and $\mathcal{L}_{\text{rec}}$.}
	\label{fig:overview} 
	\vspace{-1.2em} 
\end{figure}

\paragraph{Overview} Once trained, \tool{} is mounted as follows (\cref{fig:overview}).
As in standard FL, the client propagates their batch $(\mX, \vy)$ of $B$ examples $(\vx_i, \vy_i)$ through the shared model $f$ with parameters $\param_f$ sent by the server, and returns the gradient $\vg$ (for simplicity we assume FedSGD) of the public loss $\ell$ w.r.t. $\param_f$.
When $B>1$, $\vg$ aggregates gradients $\vg_i$ of individual examples, \ie $\vg = (1 / B) \sum_{i=1}^B \vg_i$. When \emph{secure aggregation} is used (discussed shortly), the sum also includes gradients of other clients.

The server's goal is to break this aggregation.
To this end, the server feeds $\vg$ to a \emph{secret decoder} consisting of a \emph{disaggregator} $d$, followed by a \emph{reconstructor} $r$. Crucially, $d$ is trained to project $\vg$ onto a hidden space in which the gradient projections of all images not satisfying some property $\mathcal{P}$ are removed. While the exact choice of $\mathcal{P}$ is not essential, the goal is that for most batches \emph{only one batch example satisfies $\mathcal{P}$}. In~\cref{fig:overview}, $\mathcal{P}=$``images with brightness at most $\tau$'' with $\tau$ chosen so only $\vx_5$ satisfies it, allowing $d$ to extract the projected gradient $d(\vg_5)$, and $r$ to steal the client image $\vx_5$.

To train \tool{}, the server chooses $\mathcal{P}$ and interprets $\param_f, \param_d$, and $\param_r$ as an encoder-decoder framework, trained end-to-end using auxiliary data to simulate a real client. 
The goal of training is for $d$ to nullify the contributions of images not satisfying $\mathcal{P}$ ($\mathcal{L}_{\text{nul}}$ in~\cref{fig:overview}), and for $r$ to reconstruct the image satisfying $\mathcal{P}$ from the output of $d$ ($\mathcal{L}_{\text{rec}}$).
The shared model $f$ is also trained to encode client data in the gradient space in a way that supports the goals of disaggregation and reconstruction.
 
\subsection{Key Components of \tool{}} \label{sec:attack:components}

We next describe the individual components of \tool{} in more detail.

\paragraph{Selecting the property $\mathcal{P}$}
Let $I_{\text{nul}} \subseteq [B] \coloneqq  \{1, \ldots, B\}$ denote the set of examples in the client batch that do not satisfy the secret property $\mathcal{P}$, and $I_{\text{rec}} \subseteq [B]$ the set of those that do. Following ~\citet{rtf}, we use properties of the type $m(\vx)<\tau$, where $m$ represents some image measurement, \eg brightness, and $\tau$ is chosen to maximize $P(|I_{\text{rec}}|=1)$. Similarly to \citet{rtf}, in our experiments we use $m$ based on brightness and color, but we emphasize that the attacker can use many different types of $m$ (as we experimentally demonstrate in \cref{sec:app_props}), \ie this choice is non-restrictive and is simply a way to single out images from big batches.

\citet{rtf} propose the \emph{global} setting, where the server sets a single value of $\tau$ for all client batches.
In particular, they choose $\tau$ such that the probability of satisfying $\mathcal{P}$ is $1/B$ on the distribution of $m$ over the whole space of input examples.
~\citet{fishing} later showed that this choice is optimal in the global setting, resulting in $P(|I_{\text{rec}}|=1)\to1/e$ from above as $B \to \infty$.
We improve upon this by allowing $\tau$ to be dependent on the client batch data $(\mX, \vy)$, \ie we propose a \emph{local} setting. 
We make the novel observation that when batch normalization (BN) is present (like in most convolutional networks), we can choose $\mathcal{P}$ with respect to the local (in-batch) distribution of $m$, \eg as the minimal brightness in every batch. 
We find that \tool{} can be trained on such local $\mathcal{P}$ with auxiliary data, empirically achieving $P(|I_{\text{rec}}|=1)>0.9$ for $B$ as large as $512$, which is a significant improvement over the probability of $1/e$ achieved in the global setting. 
Our method is enabled by the fact that each BN layer normalizes the distribution of its input, intertwining the computational graphs of images in the batch which are otherwise independent.
Unlike prior work, our local properties $\mathcal{P}$, allow the attacker to, most of the time, steal the client data after only a single communication round.

We note that \emph{secure aggregation}~\citep{secagg} is more challenging, and \emph{not equivalent} to a large batch from one client when BN is present, an aspect overlooked in prior work.
To overcome this, we design a more elaborate $\mathcal{P}$ that combines local and global properties, resulting in $P(|I_{\text{rec}}|=1)\to1/e$, as in prior work.
We provide a detailed explanation in \cref{sec:app_secagg}, and in our experiments in~\cref{sec:experiments} evaluate both large batch and secure aggregation variants of \tool{}.

\paragraph{Training $\param_f$ for suitable gradient encodings}
Training the shared model weights $\param_f$ along with the secret decoder  is essential for the success of \tool{}.
Intuitively, we can interpret the client-side gradient computation as a latent space encoding of the client data. 
The failures of honest attacks, discussed in~\cref{sec:related}, suggest that the gradient encoding often lacks the required information to reconstruct user data.
Our key observation is that the MS threat model uniquely allows to overcome this issue by \emph{controlling the gradient encoding} by tuning $\param_f$.
In particular, we maliciously optimize $\param_f$ with SGD to allow the recovery of a single example by the other modules of \tool{}, regardless of the information lost at the encoding step.
While \citet{zhang23} also considered optimization-based modifications with auxiliary data, their approach still inherits the fundamental limitations of all boosted analytical attacks, requiring additional handcrafted modifications which, as noted in~\cref{sec:detection}, are easily weight space detectable---an issue that \tool{} circumvents by design.

\paragraph{Training $\param_d$ for secret disaggregation}
The secret disaggregator $d$ addresses the key limitation of example disaggregation attacks (discussed in \cref{sec:detection}), \ie disaggregating examples in the gradient space. 
In contrast, $d$ embeds the gradients from ${\vg}$ into a lower-dimensional space $\mathbb{R}^{n_d}$ using a secret linear map $\param_d$, concealing the disaggregation in the original gradient space.
The benefits of using such a linear map are twofold.
First, the linear map commutes with gradient aggregation due to additivity. Second, the lower-dimensional space allows us to more easily drive the projected gradients of $I_{\text{nul}}$ to $0$, which happens when they are in or close to the null space of $\param_d$. Combining the two properties ($(i)$ and $(ii)$ in ~\cref{eq:disagg}) ideally allows us to retain only the chosen sample from the aggregated gradient $\vg$:
\begin{equation} \label{eq:disagg}
	d(\vg)=d(\sum_{i=1}^B \vg_i)\overset{(i)}{=}\sum_{i=1}^B d(\vg_i) ={\sum_{i \in I_{\text{nul}}} d(\vg_i)} + \sum_{i \in I_{\text{rec}}} d(\vg_i) \overset{(ii)}{\approx} \sum_{i \in I_{\text{rec}}} d(\vg_i).
\end{equation}
To achieve this in practice, $f$ and $d$ should be set such that $d(\vg_i)\approx 0$ for all $i \in I_{\text{nul}}$, while tolerating $d(\vg_i)\neq 0$ for the single $i \in I_{\text{rec}}$.
To this end, for $\mathcal{P}$ chosen as discussed above, we define the following objective:
\begin{equation} \label{eq:nul}
	\mathcal{L}_{\text{nul}} = \sum_{i \in I_{\text{nul}}}\|\,d(\vg_i)\,\|^2_2,
\end{equation}
which \tool{} aims to minimize during training.
We ensure that this does not also nullify $d(\vg_i)$ for the example of interest in $I_{\text{rec}}$, so $r$ is able to recover that example from $d(\vg_i)$, as described next.

\begin{wrapfigure}[22]{L}{0.55\textwidth}
	\begin{minipage}{0.55\textwidth}
\vspace{-2.4em}
\begin{algorithm}[H]
	\caption{The training procedure of \tool{}}
	\label{alg:tool}
	\begin{algorithmic}[1]
		\Function{Train\tool{}}{$f,\,\ell,\,B,\,\mathcal{X},\,\mathcal{Y}$}
		\State Choose $\mathcal{P}$, initialize $d$ and $r$
		\While{not converged}
				\State $\mX,\,\vy \leftarrow \{\vx_i, \vy_i \sim (\mathcal{X},\,\mathcal{Y})\,|\, i\in[B]\}$\label{algline:sample}
				\State $I_{\text{nul}}, I_{\text{rec}} \leftarrow \mathcal{P}(\mX,\vy) $\label{algline:property}
				\State $\mX_{\text{nul}}, \vy_{\text{nul}} \leftarrow \mX[I_{\text{nul}}], \vy[I_{\text{nul}}]$
				\State $\mX_{\text{rec}}, \vy_{\text{rec}} \leftarrow \mX[I_{\text{rec}}], \vy[I_{\text{rec}}]$
				\State $\vg_{\text{nul}},\vg_{\text{rec}}\leftarrow$\Call{BP}{$f,\ell,\mX_{\text{nul}}, \mX_{\text{rec}}, \vy_{\text{nul}}, \vy_{\text{rec}}$}\label{algline:grads}
				\State $\mathcal{L}_{\text{nul}} \leftarrow \|\,d(\vg_{\text{nul}})\,\|^2_2$ \Comment{\cref{eq:nul}}
				\State $\mathcal{L}_{\text{rec}} \leftarrow \| \, r( d(\vg_{\text{rec}})) - \mX_{\text{rec}}\,\|^2_2$ \Comment{\cref{eq:rec}}
				\State $\mathcal{L}\leftarrow \mathcal{L}_{\text{rec}} + \alpha\cdot\mathcal{L}_{\text{nul}}$\label{algline:obj} \Comment{\cref{eq:fullloss}}
				\State $\param_m\leftarrow \param_m - \gamma_m \cdot \frac{\partial\mathcal{L}}{\partial \param_m}, \forall m \in \{f,d,r\}$\label{algline:sgd}
			\EndWhile
			\State\Return $f,d, r$
		\EndFunction
		\Function{BP}{$f,\,\ell,\,\mX_{\text{nul}},\, \mX_{\text{rec}}, \,\vy_{\text{nul}}, \,\vy_{\text{rec}}$}
			\State $ \llbracket\vl_{\text{nul}};\vl_{\text{rec}}\rrbracket\leftarrow \ell( f(\llbracket \mX_{\text{nul}};\mX_{\text{rec}} \rrbracket),\llbracket \vy_{\text{nul}}; \vy_{\text{rec}} \rrbracket )$  
			\State\Return $\frac{\partial \vl_{\text{nul}}}{\partial \param_f},\frac{\partial \vl_{\text{rec}}}{\partial \param_f}$
		\EndFunction
	\end{algorithmic}
\end{algorithm}

  \end{minipage}
\end{wrapfigure}  
\paragraph{Training $\param_r$ for image reconstruction}
The final component of \tool{} we discuss is the secret reconstructor \mbox{$r\colon \mathbb{R}^{n_d}\rightarrow\mathbb{R}^{n_r}$}, which receives $d(\vg)$, \ie the (noisy) isolated embedding of the target image gradient, as seen in~\cref{eq:disagg}.
The reconstructor aims to map $d(\vg)$ back to the original image $\vx_{\text{rec}}$, effectively stealing that example from the original batch, compromising client privacy.
To this end, we define the following $\ell_2$ reconstruction objective, which is at odds with $\mathcal{L}_{\text{nul}}$:
\begin{equation} \label{eq:rec}
	\mathcal{L}_{\text{rec}} = \| \, r(d(\vg_{\text{rec}})) - \vx_{\text{rec}}\,\|^2_2.
\end{equation}
The final loss function of \tool{} weighs the two losses using a hyperparameter $\alpha>0$:
\begin{equation} \label{eq:fullloss}
	\mathcal{L} = \mathcal{L}_{\text{rec}} + \alpha\cdot\mathcal{L}_{\text{nul}}.
\end{equation}
All three key components of \tool{} are jointly trained to minimize $\mathcal{L}$.

\subsection{End-to-end Attack Description \& Discussion} \label{sec:attack:final}

\cref{alg:tool} describes the training of \tool{}.
We train on client-sized batches (see \cref{sec:app_robustness} for a related study) sampled from our auxiliary data (\cref{algline:sample}). Based on $\mathcal{P}$, we select the index sets $I_{\text{nul}}$ and $I_{\text{rec}}$ (\cref{algline:property}), representing the examples we aim to disaggregate. Then, we simulate the client updates $\bm{g}_{\text{rec}}$ and $\bm{g}_{\text{nul}}$ computed on the full batch $\bm{X}$ (\cref{algline:grads}), and use them to compute our optimization objective (\cref{algline:obj}).
We minimize the objective by jointly training $f$, $d$, and $r$ using SGD (\cref{algline:sgd}).

\begin{wrapfigure}[7]{R}{0.39\textwidth}
	\begin{minipage}{0.39\textwidth}
		\vspace{-2.1em}
		\begin{algorithm}[H] 
			\caption{Mounting \tool{}}
			\label{alg:toolattack}
			\begin{algorithmic}[1]
				\Function{Mount\tool{}}{$f,\,d,\, r$}
				\State $\vg\leftarrow $\Call{GetClientUpdate}{$f$}\label{algline:send}
				\State $\vx_{\text{stolen}}\leftarrow r(d(\vg))$\label{algline:dis_and_rec}
				\State\Return $\vx_{\text{stolen}}$
				\EndFunction
			\end{algorithmic}
		\end{algorithm}
		\vspace{-2.7em}
	\end{minipage}
\end{wrapfigure} 

Mounting \tool{} once the malicious weights $\theta_f$ have been trained using~\cref{alg:tool} is simple, as we illustrate in~\cref{alg:toolattack}.
The server, during an FL round, sends the client the malicious model $f$ (\cref{algline:send}), and receives the gradient update $\vg$.
Then, it applies its secret disaggregator $d$ and reconstructor $r$ (\cref{algline:dis_and_rec}) to obtain $\vx_{\text{stolen}}$, the reconstructed private example from the client batch.

\paragraph{\tool{} satisfies all requirements}
We now reflect on the requirements listed in~\cref{sec:detection} and discuss how \tool{} satisfies them.
First, \tool{} does not utilize any attack vector apart from maliciously modifying the weights of $f$, does not assume unrealistic knowledge of BN statistics or batch labels, and makes no assumptions regarding label distributions, in contrast with some prior work~\citep{nvidia,fishing}.
We remark that the necessity of such side information is the artifact of optimization-based attacks, and another reason why approaches that do not attempt to lift honest attacks (such as \tool{}) may be more promising.
\tool{} was greatly influenced by the assumption that clients \emph{will} inspect the models, aiming to detect malicious updates.
Namely, \tool{} avoids weight space detectable handcrafted modifications and introduces secret disaggregation as means to also avoid gradient space detection.
As we show in~\cref{sec:experiments}, \tool{} successfully steals client data on realistic convolutional networks with large batch sizes and secure aggregation, demonstrating its practicality. 
\section{Experimental Evaluation} \label{sec:experiments} 
In this section, we present our experimental results, demonstrating that \tool{} is effective at reconstructing client images from realistic networks, in both large batch and secure aggregation settings.
These results are especially valuable given the important advantages of \tool{} over prior work in terms of satisfying the requirements for practical attacks (\cref{sec:detection}), as we have discussed in~\cref{sec:attack}.

\paragraph{Experimental setup}
We use ResNet18~\citep{resnet} in all experiments. 
We use the \emph{CIFAR10} dataset, as well as \emph{CIFAR100}~\citep{cifar10} and \emph{ImageNet} \citep{imagenet}, to demonstrate the ability of \tool{} to scale with the number of labels and input size, respectively. 
We generally use the training set as auxiliary data, and mount the attack on randomly sampled batches of size $B$ from the test set for CIFAR10/100 and validation set for ImageNet. We further experiment with auxiliary datasets of different sizes in \cref{sec:app_size_data}, and clients with different heterogeneity levels in \cref{app:hetero} where we show that \tool{} is highly effective even when only small amount of auxiliary data is available and when clients data is highly heterogeneous. We run all experiments on a single NVIDIA A100 GPU with $40$GB (CIFAR10/100) and $80$GB (ImageNet) of VRAM.
Each CIFAR experiment took $<7$ GPU days to train and $<1$h to mount on $1000$ batches. 
The ImageNet model trained for $14$ GPU days, with $0.5h$ to mount the attack on $100$ batches.
In our CIFAR experiments, we set $r$ to a linear layer and subsume $d$ in it. For ImageNet, we use a linearized U-Net decoder~\citep{unet} (see \cref{sec:app_unet}).
We defer additional implementation details to \cref{sec:app_hyperparams} and \cref{sec:app_imp_detail}. 

In all experiments, we use the properties of maximal brightness (\emph{Bright}) and redness (\emph{Red}), training separate malicious weights for each (dataset, property, batch size) triple.
We report 3 reconstruction quality metrics:
(i) the fraction of good reconstructions (\emph{Rec}), \ie batches where reconstructions have PSNR $> 19$~\citep{psnrvsssim} to the ground truth;
(ii) the average PSNR across all attacked batches (\emph{PSNR-All}); and 
(iii) the average PSNR for the top $\frac{1}{e}\approx37\%$ of the batch reconstructions (\emph{PSNR-Top}) that allows to compare \tool{} in large batch (1 client) and secure aggregation (many clients) settings. We provide experiments with more properties and metrics in \cref{sec:app_props} and \cref{sec:app_ssim}.

\paragraph{Large batch reconstruction on CIFAR10/100}
\begin{table*}[t]\centering
	\vspace{-3em}
	\caption{Large batch reconstruction for different batch sizes $B$. The metrics are introduced at the top of~\cref{sec:experiments}. Results with 2 more settings (CIFAR100, Bright and CIFAR10, Red) are given in~\cref{sec:app_full_exp}.} \label{table:single}
	 
	\newcommand{\threecol}[1]{\multicolumn{3}{c}{#1}}
	\newcommand{\fivecol}[1]{\multicolumn{5}{c}{#1}}
	\newcommand{\ninecol}[1]{\multicolumn{9}{c}{#1}}
	
	\newcommand{\bsz}{Batch Size~}
	\newcommand{\certified}{{CR(\%)}}
	 
	\renewcommand{\arraystretch}{1.2}
	
	\newcommand{\ccellt}[2]{\colorbox{#1}{\makebox(20,8){{#2}}}}
	\newcommand{\ccellc}[2]{\colorbox{#1}{\makebox(8,8){{#2}}}}
	\newcommand{\ccells}[2]{\colorbox{#1}{\makebox(55,8){{#2}}}}
	
	\newcommand{\temp}[1]{\textcolor{red}{#1}}
	\newcommand{\noopcite}[1]{} 
	
	\newcommand{\skiplen}{-0.4em} 
	\newcommand{\rlen}{0.01\linewidth} 
	\vspace{-0.4em}
	\resizebox{0.8 \linewidth}{!}{ 
		\begingroup
		\setlength{\tabcolsep}{5pt} %
		\begin{tabular}{@{}l rrr p{\skiplen}  rrr@{}} \toprule

			& \threecol{CIFAR10, Bright} && \threecol{CIFAR100, Red}\\
			
			\cmidrule(l{5pt}r{5pt}){2-4} \cmidrule(l{5pt}r{5pt}){6-8}
			
			$B$&  Rec (\%) & PSNR-Top $\uparrow$ & PSNR-All $\uparrow$ &&  Rec (\%) & PSNR-Top $\uparrow$ & PSNR-All $\uparrow$\\ \midrule
			64 & $89.4$ & $32.1\pm2.0$ & $27.2\pm5.3$ &&  $97.1$ & $31.7\pm1.1$ & $29.0\pm3.4$\\ 
			128 & $\bm{94.2}$ & $31.9\pm1.7$ & $28.2\pm4.3$ && $97.4$ & $31.8\pm1.1$ & $29.3\pm3.2$\\
			256 & $93.5$ & $\bm{32.8\pm2.0}$ & $\bm{28.5\pm5.0}$ && $97.7$ & $31.3\pm1.0$ & $28.6\pm3.2$\\
			512 & $87.8$ & $26.6\pm1.8$ & $23.2\pm3.5$ &&
			 $\bm{98.6}$ & $\bm{33.1\pm1.1}$ & $\bm{30.5\pm3.1}$ \\
			\bottomrule
		\end{tabular}
		\endgroup
	}
\vspace{-1em}
\end{table*}
    
\begin{figure}[t]
	\centering
	\vspace{-3em}
	\includegraphics[width=0.87\linewidth]{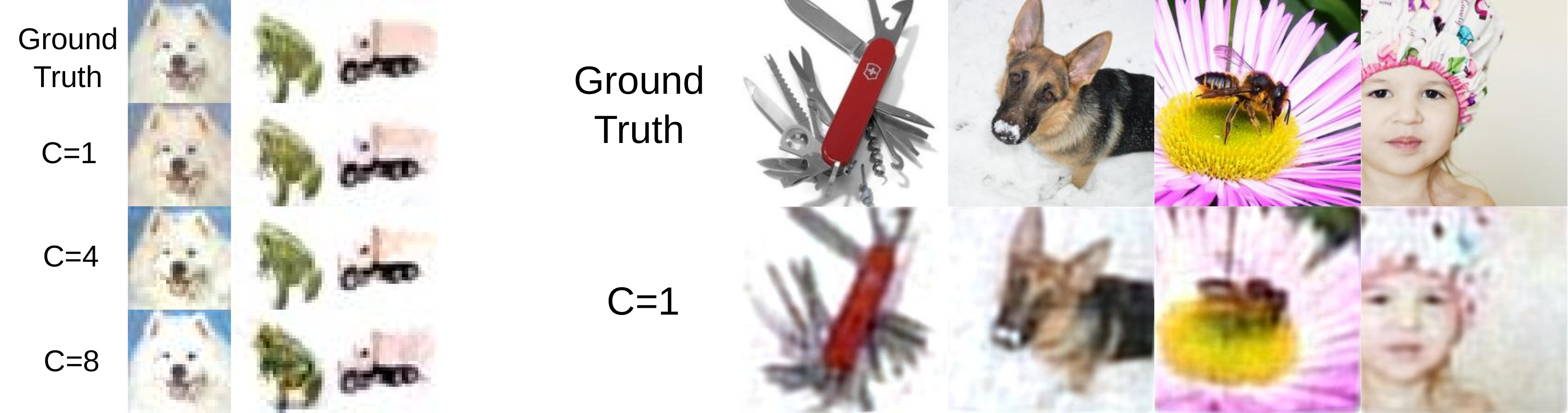}
	\caption{Example reconstructions of \tool{} with 128 total examples and different number of clients $C$ on CIFAR10~(Left) and 64 examples on ImageNet~(Right), both using the Bright property.}
	\label{fig:quality}
	\vspace{-1.8em}
\end{figure}

A subset of our main results is shown in \cref{table:single}; the full results are deferred to~\cref{sec:app_full_exp} and follow similar trends.
We make several key observations.
First, in most experiments, the use of local properties (see \cref{sec:attack:components}) allows us to steal an image from most batches (up to $98.6\%$), greatly improving over $1/e \%$ achieved by prior work.
Second, we obtain good reconstructions for both Red and Bright property (average PSNR up to $30$), which confirms that \tool{} can handle a diverse set of properties, and that property choice is not crucial for its success.
Finally, \tool{} successfully steals images even from very large batch sizes such as $512$, showing no clear degradation in performance.
On top of these quantitative results, we show example reconstructions in \cref{fig:quality} (left, $C=1$), visually confirming their quality.

\newpage

\paragraph{Large batch reconstruction on ImageNet}
To show scalability to high-res images, we train \tool{} on ImageNet with $B=64$ and the Bright property (due to high computational costs, we leave more thorough studies of ImageNet to future work).
To speed up convergence, we pretrain parts of the transposed convolution stack alongside $\theta_f$ on downsized images, and use it as initialization. We obtain PSNR-All of $21.9\pm3.5$ and PSNR-Top of $25.1\pm2.5$, corresponding to $82.1\%$ successfully attacked batches. Our results are significant, as these PSNR values on Imagenet represent very high quality reconstructions and are higher than the state-of-the-art attack in \cite{fishing}. This is further visually confirmed by the example reconstructions we show in \cref{fig:quality} (right). From the recovered images we conclude that \tool{} can be efficiently instantiated on high-resolution images, resulting in very detailed reconstructions that allow the identification of complex objects and individual people, constituting a serious violation of privacy. We note the significance of these results, as stealing even a single ImageNet image is impossible with honest attacks without restrictive assumptions~\citep{arora}. We see these results as an encouraging signal for general applicability of \tool{}.

\paragraph{Secure aggregation}
\begin{wrapfigure}[12]{R}{0.55\textwidth}
\vspace{-1.05em} 
\makeatletter
\def\@captype{table}
\makeatother
\centering
\caption{CIFAR10 reconstruction with secure aggregation, varying the number of clients ($C$) and total images (\#Imgs). See~\cref{sec:app_full_aggr_exp} for results with another property.} \label{table:secure}
\vspace{-0.4em}
			
			\newcommand{\threecol}[1]{\multicolumn{2}{c}{#1}}
			\newcommand{\fivecol}[1]{\multicolumn{5}{c}{#1}}
			\newcommand{\ninecol}[1]{\multicolumn{9}{c}{#1}}
			 
			\newcommand{\bsz}{Batch Size~}
			\newcommand{\certified}{{CR(\%)}}
			
			\renewcommand{\arraystretch}{1.2}
			
			\newcommand{\ccellt}[2]{\colorbox{#1}{\makebox(20,8){{#2}}}}
			\newcommand{\ccellc}[2]{\colorbox{#1}{\makebox(8,8){{#2}}}}
			\newcommand{\ccells}[2]{\colorbox{#1}{\makebox(55,8){{#2}}}}
			
			\newcommand{\temp}[1]{\textcolor{red}{#1}}
			\newcommand{\noopcite}[1]{} 
			 
			\newcommand{\skiplen}{-0.1em} 
			\newcommand{\rlen}{0.01\linewidth} 
			
			\resizebox{\linewidth}{!}{
				\begingroup
				\setlength{\tabcolsep}{5pt} %
				\begin{tabular}{@{}l rr p{\skiplen}  rr p{\skiplen} rr p{\skiplen}  rr p{\skiplen}  rr  p{\skiplen}  rr@{}} \toprule

					&\threecol{$C=4$, Bright} && \threecol{$C=8$, Bright}\\
					
					\cmidrule(l{5pt}r{5pt}){2-3} \cmidrule(l{5pt}r{5pt}){5-6}
					
					\#Imgs&  Rec (\%) & PSNR-Top $\uparrow$ && Rec (\%) & PSNR-Top $\uparrow$\\ \midrule
					64 & $41.4$ & $\bm{27.3\pm3.1}$ && $41.3$ & $26.6\pm3.7$ \\
					128 & $44.2$ & $26.8\pm3.0$ && $40.6$ & $\bm{27.3\pm3.3}$ \\
					256 & $51.9$ & $\bm{27.3\pm2.5}$ && $41.9$ & $25.4\pm3.1$ \\
					512 & $\bm{52.9}$ & $25.7\pm2.4$ && $\bm{51.7}$ & $25.9\pm2.8$ \\

					\bottomrule
				\end{tabular}
				\endgroup
			}
\end{wrapfigure} 
In~\cref{table:secure}, we present the results of CIFAR10 experiments with secure aggregation with $C=4$ and $C=8$ clients and batch sizes chosen to match the total number of images (\emph{\#Imgs}) as in \cref{table:single}.
Most importantly, as before, \tool{} consistently obtains image reconstructions with average \mbox{PSNR $>25$}, \ie recovers most images almost perfectly.
Comparing to \cref{table:single}, the success probability degrades with $C$, confirming our intuition (see~\cref{sec:attack}) that secure aggregation provides additional protection in the presence of BN, compared to simply using large batches. 
Despite this, the success probability Rec is significantly higher than $1/e \%$ of prior work.
We suspect this is due to the model learning a restricted version of single-client reconstruction for each client, and further compare the two variants in \cref{sec:prob_vs_rel}.
We note that Rec rises with the number of images, which we believe is due to the better estimation of the property threshold for larger batches. 
Finally, in \cref{fig:quality} (left), we can visually compare results for different number of clients, noting no obvious degradation, which reaffirms that \tool{} can breach privacy even when secure aggregation is used.

\paragraph{Robustness to distribution shifts}
\begin{wrapfigure}[11]{R}{0.52\textwidth}
	\makeatletter
	\def\@captype{table}
	\makeatother
	\vspace{-1.2em}
	\caption{\tool{} is robust to distribution shifts between the auxiliary dataset (CIFAR10) and the client dataset $D_c$. We use $B=128$ and the Red property.} \label{table:shifts}
	\vspace{-0.8em}
	\newcommand{\threecol}[1]{\multicolumn{2}{c}{#1}}
	\newcommand{\fivecol}[1]{\multicolumn{5}{c}{#1}}
	\newcommand{\ninecol}[1]{\multicolumn{9}{c}{#1}}
	  
	\newcommand{\bsz}{Batch Size~}
	\newcommand{\certified}{{CR(\%)}}
	
	\renewcommand{\arraystretch}{1.2}
	
	\newcommand{\ccellt}[2]{\colorbox{#1}{\makebox(20,8){{#2}}}}
	\newcommand{\ccellc}[2]{\colorbox{#1}{\makebox(8,8){{#2}}}}
	\newcommand{\ccells}[2]{\colorbox{#1}{\makebox(55,8){{#2}}}}
	
	\newcommand{\temp}[1]{\textcolor{red}{#1}}
	\newcommand{\noopcite}[1]{} 
	
	\newcommand{\skiplen}{0.004\linewidth} 
	\newcommand{\rlen}{0.01\linewidth} 
	
	\resizebox{\linewidth}{!}{
	\begingroup
	\setlength{\tabcolsep}{5pt}
	\begin{tabular}{@{}l rrrr@{}} \toprule
		
		$D_c$ & Rec (\%) & PSNR-Top $\uparrow$ & PSNR-All $\uparrow$\\ \midrule
		CIFAR10      & $93.5$ & $31.1\pm 1.2$ & $27.8\pm 4.1$ \\
		CIFAR10.1v6    & $\bm{96.0}$ & $\bm{31.6\pm 1.0}$ & $\bm{28.4\pm 3.8}$ \\
		CIFAR10.2    & $90.2$ & $31.6\pm 1.3$ & $27.5\pm 5.0$ \\
		TinyImageNet & $80.2$ & $27.6\pm 1.0$ & $23.7\pm 4.7$ \\
		ISIC2019 & $98.0$ & $29.4\pm 1.0$ & $26.9 \pm 2.84$ \\	
		\bottomrule
	\end{tabular}
	\endgroup
}
\end{wrapfigure}

A question that naturally arises is if the need for auxiliary data restricts the applicability of \tool{}.
In this experiment we show otherwise, demonstrating robustness to distribution shifts between the attacker's auxiliary dataset ($D_a$) and the client dataset ($D_c$), \ie an attacker can successfully mount \tool{} without the knowledge of $D_c$, relying only on public data. 
We set $C=1, B=128$, and $D_a=\text{CIFAR10}$ and explore several options for $D_c$, illustrating different levels of shift. 
Namely, CIFAR10.1v6~\citep{cifar101} and CIFAR10.2~\citep{cifar102} represent naturally occurring shifts of the data source, TinyImageNet~\citep{tinyimagenet} (mapped to 10 classes) models different data sources for $D_a$ and $D_c$, and ISIC2019~\citep{tschandl2018ham10000,8363547,combalia2019bcn20000}  models a more severe domain shift between $D_a$ and $D_c$.

The results are shown in~\cref{table:shifts}. 
We observe no degradation for CIFAR10.1v6 and CIFAR10.2, confirming that \tool{} can handle naturally-occurring data shifts. For TinyImageNet and ISIC2019, despite the large discrepancy to CIFAR10 in image and label distributions, we observe high quality reconstruction on 80\% and 98\% of images, confirming that \tool{} is not limited by the choice of $D_a$.
We further investigate the robustness to batch size mismatch in \cref{sec:app_robustness} and corruptions in \cref{app:corruptions}.
\paragraph{Comparison to prior MS attacks}
\begin{table}[t]\centering
	\vspace{-3.5em}	
	\caption{Comparison between \tool{} and prior state-of-the-art MS attacks.} \label{table:fishing_comp}
	\vspace{-0.8em}
	\newcommand{\threecol}[1]{\multicolumn{2}{c}{#1}}
	\newcommand{\fivecol}[1]{\multicolumn{5}{c}{#1}}
	\newcommand{\ninecol}[1]{\multicolumn{9}{c}{#1}}
	
	\newcommand{\bsz}{Batch Size~}
	\newcommand{\certified}{{CR(\%)}}
	
	\renewcommand{\arraystretch}{1.2}
	
	\newcommand{\ccellt}[2]{\colorbox{#1}{\makebox(20,8){{#2}}}}
	\newcommand{\ccellc}[2]{\colorbox{#1}{\makebox(8,8){{#2}}}}
	\newcommand{\ccells}[2]{\colorbox{#1}{\makebox(55,8){{#2}}}}
	
	\newcommand{\temp}[1]{\textcolor{red}{#1}}
	\newcommand{\noopcite}[1]{} 
	
	\newcommand{\skiplen}{0.004\linewidth} 
	\newcommand{\rlen}{0.01\linewidth} 
	 
	\resizebox{0.85\linewidth}{!}{
	\begingroup
	\setlength{\tabcolsep}{5pt}
	\begin{tabular}{@{}lrrrrr@{}} \toprule
	Method & Und-Rec (\%) & PSNR-Und-Rec $\uparrow$ & PSNR-Und $\uparrow$ & Rec (\%) & PSNR-All $\uparrow$ \\\midrule
	Fishing $\beta=400$ & $4$ & $20.2\pm 0.5$ & $17.1\pm 2.3$& $77$ & $21.7\pm 3.2$ \\
	Fishing $\beta=100$ & $8$ & $20.6\pm 1.6$ & $16.4\pm 2.9$& $63$ & $20.2\pm 4.1$ \\	
	Fishing $\beta=50$ & $4$ & $19.4\pm 0.3$ & $15.5\pm 2.2$ & $52$ & $19.4\pm 4.5$ \\	
	Fishing $\beta=12.5$ & $1$ & $22.4\pm 0.0$ & $13.7\pm 2.0$& $8$ & $14.5\pm 3.4$ \\	
	Zhang23 & $0$ & $N/A$ & $N/A$& $5$ & $15.8\pm 1.8$ \\	
	LOKI & $0$ & $N/A$ & $N/A$& $\bm{100}$ & $\bm{143.4\pm 10.3}$ \\	
	\tool{} & $\bm{90}$ & $\bm{24.6\pm 2.2}$ & $\bm{23.8\pm 3.3}$& $90$ & $23.8\pm 3.3$ \\
	\bottomrule
\end{tabular}

	\endgroup
}
\vspace{-1.6em}
\end{table}

We compare \tool{} to 3 state-of-the-art MS attacks: (i) \emph{Fishing}~\citep{fishing}, an example disaggregation attack; (ii) \emph{Zhang23}~\citep{zhang23}, a boosted analytical attack; and (iii) \emph{LOKI}~\citep{mandrake}, a boosted analytical attack that relies on a stronger threat model that permits architectural changes and sending different models to clients. 
We attack 100 batches on CIFAR10, with $C=1, B=128$. We use the Red property for \tool{}.
As in~\cref{sec:detection}, we explore different variants of Fishing by varying the parameter $\beta$ which should control the strength-detectability tradeoff.
We report the usual metrics \emph{Rec} and \emph{PSNR-All}, the percentage of undetected successful attacks (\emph{Und-Rec}) based on D-SNR (\cref{sec:detection}) and T-SNR (\cref{sec:app_detection}), as well as the average PSNR of all undetected reconstructions (\emph{PSNR-Und}), and the successful undetected reconstructions (\emph{PSNR-Und-Rec}). We provide more details about the experimental setup in \cref{sec:app_fishing_comp_details}.

The results are shown in \cref{table:fishing_comp}.
Setting detectability aside, \tool{} outperforms all methods but LOKI, while also being very fast to mount (<2 sec per batch). We emphasize that LOKI's performance is largely due to its architectural changes to the ResNet which are trivially detectable by clients and crucial to the application of the method. We also observe that Zhang23 fails to recover most images at all due to the stride>1 and pooling in realistic networks that cause severe downscaling of the image fed to the attacked linear layer as discussed in \cref{sec:detection}.
Crucially, only a tiny fraction of successful Fishing attacks are undetected, while other prior methods completely fail to avoid detection. In contrast, for \tool{} \emph{all} successful attacks remain undetected.
Finally, we confirm our observation from \cref{fig:dsnr} that prior MS attacks need to jeopardize reconstruction quality to avoid detection, as for Fishing {PSNR-Und} is well below {PSNR-All} for all values of $\beta$.
Our experiments reaffirm that the reliance on honest attacks of prior MS attacks makes them easily detectable, and thus unrealistic.

\section{Outlook} \label{sec:mitigations}

While \tool{} is a powerful attack that can harm user privacy, we believe our work opens the door to a more principled investigation of defenses, as it illustrates that techniques such as secure aggregation are not as effective as previously thought. To mitigate attacks like ours, prior work has discussed cryptographic techniques like SMPC or FHE, which are still largely impractical~\citep{advances}, as well as differential privacy methods, which we demonstrate in \cref{app:defenses} to not be effective enough.

Thus, we believe that principled client-side detection is the most promising way forward.
While \tool{} avoids pitfalls of prior attacks which make them easily detectable, and we see no clear ways to detect it currently, more mature detection techniques may be able to do so.
We encourage such work and advocate for efficient and robust checks based on attack categorization (such as in this work), as opposed to ad-hoc detection which attacks can easily adapt to.
On the attack side, interesting future directions include other data modalities and model architectures and improving \tool{}'s training cost.
\section{Conclusion} \label{sec:conclusion}
In this work, we explored the issue of client-side detectability of malicious server (MS) attacks in federated learning.
We demonstrated that all prior attacks are detectable in a principled way, and proposed \tool{}, a novel attack strategy that by design avoids such detection while effectively stealing data despite aggregation. 
Our work highlights the importance of studying attack detectability and represents a promising first step towards MS attacks that compromise privacy in realistic settings.     
\clearpage
\subsubsection*{Ethics Statement}
As we noted in~\cref{sec:mitigations}, the attack introduced by this work, \tool{}, advances the capabilities of attackers aiming to compromise client privacy in FL.
Further, as~\citet{fishing} point out, attacks like ours based on property thresholding can lead to disparate impact, affecting outlier groups more severely as their inputs are more likely to be reconstructed.
However, we believe that our principled investigation of detection and the emphasis on realistic scenarios, as well as making the details of our attack public and open source (which we intend to do after publication), both have a significant positive impact, as they open the door to further systematic studies of defenses, and help practitioners better estimate the privacy risks of their FL deployments and avoid the common error of underestimating the vulnerability.

\subsubsection*{Acknowledgments}
This research was partially funded by the Ministry of Education and Science of Bulgaria (support for INSAIT, part of the Bulgarian National Roadmap for Research Infrastructure).

\message{^^JLASTBODYPAGE \thepage^^J}

\bibliography{references}

\begin{thebibliography}{51}
\providecommand{\natexlab}[1]{#1}
\providecommand{\url}[1]{\texttt{#1}}
\expandafter\ifx\csname urlstyle\endcsname\relax
  \providecommand{\doi}[1]{doi: #1}\else
  \providecommand{\doi}{doi: \begingroup \urlstyle{rm}\Url}\fi

\bibitem[Abadi et~al.(2016)Abadi, Chu, Goodfellow, McMahan, Mironov, Talwar,
  and Zhang]{dpsgd}
Mart{\'{\i}}n Abadi, Andy Chu, Ian~J. Goodfellow, H.~Brendan McMahan, Ilya
  Mironov, Kunal Talwar, and Li~Zhang.
\newblock Deep learning with differential privacy.
\newblock In \emph{{CCS}}, 2016.

\bibitem[Balunovic et~al.(2022{\natexlab{a}})Balunovic, Dimitrov, Jovanovic,
  and Vechev]{LAMP}
Mislav Balunovic, Dimitar~I. Dimitrov, Nikola Jovanovic, and Martin~T. Vechev.
\newblock {LAMP:} extracting text from gradients with language model priors.
\newblock In \emph{NeurIPS}, 2022{\natexlab{a}}.

\bibitem[Balunovic et~al.(2022{\natexlab{b}})Balunovic, Dimitrov, Staab, and
  Vechev]{bayesian}
Mislav Balunovic, Dimitar~Iliev Dimitrov, Robin Staab, and Martin~T. Vechev.
\newblock Bayesian framework for gradient leakage.
\newblock In \emph{{ICLR}}, 2022{\natexlab{b}}.

\bibitem[Boenisch et~al.(2021)Boenisch, Dziedzic, Schuster, Shamsabadi,
  Shumailov, and Papernot]{cah}
Franziska Boenisch, Adam Dziedzic, Roei Schuster, Ali~Shahin Shamsabadi, Ilia
  Shumailov, and Nicolas Papernot.
\newblock When the curious abandon honesty: Federated learning is not private.
\newblock \emph{arXiv}, 2021.

\bibitem[Boenisch et~al.(2023)Boenisch, Dziedzic, Schuster, Shamsabadi,
  Shumailov, and Papernot]{boenisch23}
Franziska Boenisch, Adam Dziedzic, Roei Schuster, Ali~Shahin Shamsabadi, Ilia
  Shumailov, and Nicolas Papernot.
\newblock Reconstructing individual data points in federated learning hardened
  with differential privacy and secure aggregation.
\newblock \emph{arXiv}, 2023.

\bibitem[Bonawitz et~al.(2016)Bonawitz, Ivanov, Kreuter, Marcedone, McMahan,
  Patel, Ramage, Segal, and Seth]{secagg}
Kallista~A. Bonawitz, Vladimir Ivanov, Ben Kreuter, Antonio Marcedone,
  H.~Brendan McMahan, Sarvar Patel, Daniel Ramage, Aaron Segal, and Karn Seth.
\newblock Practical secure aggregation for federated learning on user-held
  data.
\newblock \emph{NIPS}, 2016.

\bibitem[Chu et~al.(2023)Chu, Geiping, Fowl, Goldblum, and Goldstein]{panning}
Hong-Min Chu, Jonas Geiping, Liam~H Fowl, Micah Goldblum, and Tom Goldstein.
\newblock Panning for gold in federated learning: Targeted text extraction
  under arbitrarily large-scale aggregation.
\newblock \emph{ICLR}, 2023.

\bibitem[Codella et~al.(2018)Codella, Gutman, Celebi, Helba, Marchetti, Dusza,
  Kalloo, Liopyris, Mishra, Kittler, and Halpern]{8363547}
Noel C.~F. Codella, David Gutman, M.~Emre Celebi, Brian Helba, Michael~A.
  Marchetti, Stephen~W. Dusza, Aadi Kalloo, Konstantinos Liopyris, Nabin
  Mishra, Harald Kittler, and Allan Halpern.
\newblock Skin lesion analysis toward melanoma detection: A challenge at the
  2017 international symposium on biomedical imaging (isbi), hosted by the
  international skin imaging collaboration (isic).
\newblock In \emph{2018 IEEE 15th International Symposium on Biomedical Imaging
  (ISBI 2018)}, pp.\  168--172, 2018.
\newblock \doi{10.1109/ISBI.2018.8363547}.

\bibitem[Combalia et~al.(2019)Combalia, Codella, Rotemberg, Helba, Vilaplana,
  Reiter, Carrera, Barreiro, Halpern, Puig, et~al.]{combalia2019bcn20000}
Marc Combalia, Noel~CF Codella, Veronica Rotemberg, Brian Helba, Veronica
  Vilaplana, Ofer Reiter, Cristina Carrera, Alicia Barreiro, Allan~C Halpern,
  Susana Puig, et~al.
\newblock Bcn20000: Dermoscopic lesions in the wild.
\newblock \emph{arXiv preprint arXiv:1908.02288}, 2019.

\bibitem[Deng et~al.(2009)Deng, Dong, Socher, Li, Li, and Fei-Fei]{imagenet}
Jia Deng, Wei Dong, Richard Socher, Li-Jia Li, Kai Li, and Li~Fei-Fei.
\newblock Imagenet: A large-scale hierarchical image database.
\newblock In \emph{2009 IEEE conference on computer vision and pattern
  recognition}, pp.\  248--255. Ieee, 2009.

\bibitem[FedAI()]{webank}
FedAI.
\newblock Utilization of fate in risk management of credit in small and micro
  enterprises.
\newblock
  \href{https://www.fedai.org/casesutilization-of-fate-in-risk-management-of-credit-in-small-and-micro-enterprises/}{https://www.fedai.org/}.

\bibitem[Fowl et~al.(2022{\natexlab{a}})Fowl, Geiping, Reich, Wen, Czaja,
  Goldblum, and Goldstein]{decepticons}
Liam Fowl, Jonas Geiping, Steven Reich, Yuxin Wen, Wojtek Czaja, Micah
  Goldblum, and Tom Goldstein.
\newblock Decepticons: Corrupted transformers breach privacy in federated
  learning for language models.
\newblock \emph{ICLR}, 2022{\natexlab{a}}.

\bibitem[Fowl et~al.(2022{\natexlab{b}})Fowl, Geiping, Czaja, Goldblum, and
  Goldstein]{rtf}
Liam~H. Fowl, Jonas Geiping, Wojciech Czaja, Micah Goldblum, and Tom Goldstein.
\newblock Robbing the fed: Directly obtaining private data in federated
  learning with modified models.
\newblock In \emph{{ICLR}}, 2022{\natexlab{b}}.

\bibitem[Fung et~al.(2020)Fung, Yoon, and Beschastnikh]{sybils}
Clement Fung, Chris J.~M. Yoon, and Ivan Beschastnikh.
\newblock The limitations of federated learning in sybil settings.
\newblock In \emph{{RAID}}, 2020.

\bibitem[Geiping et~al.(2020)Geiping, Bauermeister, Dr{\"o}ge, and
  Moeller]{geiping}
Jonas Geiping, Hartmut Bauermeister, Hannah Dr{\"o}ge, and Michael Moeller.
\newblock Inverting gradients-how easy is it to break privacy in federated
  learning?
\newblock \emph{NeurIPS}, 2020.

\bibitem[Geng et~al.(2021)Geng, Mou, Li, Li, Beyan, Decker, and Rong]{aaai}
Jiahui Geng, Yongli Mou, Feifei Li, Qing Li, Oya Beyan, Stefan Decker, and
  Chunming Rong.
\newblock Towards general deep leakage in federated learning.
\newblock \emph{arXiv}, 2021.

\bibitem[Gupta et~al.(2022)Gupta, Huang, Zhong, Gao, Li, and Chen]{princeton}
Samyak Gupta, Yangsibo Huang, Zexuan Zhong, Tianyu Gao, Kai Li, and Danqi Chen.
\newblock Recovering private text in federated learning of language models.
\newblock In \emph{NeurIPS}, 2022.

\bibitem[He et~al.(2016)He, Zhang, Ren, and Sun]{resnet}
Kaiming He, Xiangyu Zhang, Shaoqing Ren, and Jian Sun.
\newblock Deep residual learning for image recognition.
\newblock In \emph{Proceedings of the IEEE conference on computer vision and
  pattern recognition}, pp.\  770--778, 2016.

\bibitem[Hendrycks \& Dietterich(2018)Hendrycks and
  Dietterich]{common_corruptions}
Dan Hendrycks and Thomas~G Dietterich.
\newblock Benchmarking neural network robustness to common corruptions and
  surface variations.
\newblock \emph{arXiv preprint arXiv:1807.01697}, 2018.

\bibitem[Horé \& Ziou(2010)Horé and Ziou]{psnrvsssim}
Alain Horé and Djemel Ziou.
\newblock Image quality metrics: Psnr vs. ssim.
\newblock In \emph{2010 20th International Conference on Pattern Recognition},
  pp.\  2366--2369, 2010.
\newblock \doi{10.1109/ICPR.2010.579}.

\bibitem[Huang et~al.(2021)Huang, Gupta, Song, Li, and Arora]{arora}
Yangsibo Huang, Samyak Gupta, Zhao Song, Kai Li, and Sanjeev Arora.
\newblock Evaluating gradient inversion attacks and defenses in federated
  learning.
\newblock In \emph{NeurIPS}, 2021.

\bibitem[Jin et~al.(2021)Jin, Chen, Hsu, Yu, and Chen]{cafe}
Xiao Jin, Pin{-}Yu Chen, Chia{-}Yi Hsu, Chia{-}Mu Yu, and Tianyi Chen.
\newblock {CAFE:} catastrophic data leakage in vertical federated learning.
\newblock \emph{arXiv}, 2021.

\bibitem[Kairouz et~al.(2019)Kairouz, McMahan, Avent, Bellet, Bennis, Bhagoji,
  Bonawitz, Charles, Cormode, Cummings, D'Oliveira, Rouayheb, Evans, Gardner,
  Garrett, Gasc{\'{o}}n, Ghazi, Gibbons, Gruteser, Harchaoui, He, He, Huo,
  Hutchinson, Hsu, Jaggi, Javidi, Joshi, Khodak, Kone{\v{c}}n{\'y}, Korolova,
  Koushanfar, Koyejo, Lepoint, Liu, Mittal, Mohri, Nock, {\"{O}}zg{\"{u}}r,
  Pagh, Raykova, Qi, Ramage, Raskar, Song, Song, Stich, Sun, Suresh,
  Tram{\`{e}}r, Vepakomma, Wang, Xiong, Xu, Yang, Yu, Yu, and Zhao]{advances}
Peter Kairouz, H.~Brendan McMahan, Brendan Avent, Aur{\'{e}}lien Bellet, Mehdi
  Bennis, Arjun~Nitin Bhagoji, Kallista~A. Bonawitz, Zachary Charles, Graham
  Cormode, Rachel Cummings, Rafael G.~L. D'Oliveira, Salim~El Rouayheb, David
  Evans, Josh Gardner, Zachary Garrett, Adri{\`{a}} Gasc{\'{o}}n, Badih Ghazi,
  Phillip~B. Gibbons, Marco Gruteser, Za{\"{\i}}d Harchaoui, Chaoyang He, Lie
  He, Zhouyuan Huo, Ben Hutchinson, Justin Hsu, Martin Jaggi, Tara Javidi,
  Gauri Joshi, Mikhail Khodak, Jakub Kone{\v{c}}n{\'y}, Aleksandra Korolova,
  Farinaz Koushanfar, Sanmi Koyejo, Tancr{\`{e}}de Lepoint, Yang Liu, Prateek
  Mittal, Mehryar Mohri, Richard Nock, Ayfer {\"{O}}zg{\"{u}}r, Rasmus Pagh,
  Mariana Raykova, Hang Qi, Daniel Ramage, Ramesh Raskar, Dawn Song, Weikang
  Song, Sebastian~U. Stich, Ziteng Sun, Ananda~Theertha Suresh, Florian
  Tram{\`{e}}r, Praneeth Vepakomma, Jianyu Wang, Li~Xiong, Zheng Xu, Qiang
  Yang, Felix~X. Yu, Han Yu, and Sen Zhao.
\newblock Advances and open problems in federated learning.
\newblock \emph{arXiv}, 2019.

\bibitem[Kariyappa et~al.(2022)Kariyappa, Guo, Maeng, Xiong, Suh, Qureshi, and
  Lee]{cpa}
Sanjay Kariyappa, Chuan Guo, Kiwan Maeng, Wenjie Xiong, G.~Edward Suh,
  Moinuddin~K. Qureshi, and Hsien{-}Hsin~S. Lee.
\newblock Cocktail party attack: Breaking aggregation-based privacy in
  federated learning using independent component analysis.
\newblock \emph{arXiv}, 2022.

\bibitem[Krizhevsky et~al.(2009)Krizhevsky, Hinton, et~al.]{cifar10}
Alex Krizhevsky, Geoffrey Hinton, et~al.
\newblock Learning multiple layers of features from tiny images.
\newblock 2009.

\bibitem[Lam et~al.(2021)Lam, Wei, Brooks, Reddi, and Mitzenmacher]{lam}
Maximilian Lam, Gu{-}Yeon Wei, David Brooks, Vijay~Janapa Reddi, and Michael
  Mitzenmacher.
\newblock Gradient disaggregation: Breaking privacy in federated learning by
  reconstructing the user participant matrix.
\newblock In \emph{{ICML}}, 2021.

\bibitem[Lauer(2017)]{vc}
F.~Lauer.
\newblock Vc-dimension of hyperplanes.
\newblock In \emph{An interactive journey into machine learning}. 2017.
\newblock URL \url{https://mlweb.loria.fr/book/en/VCdimhyperplane.html}.

\bibitem[Le \& Yang(2015)Le and Yang]{tinyimagenet}
Ya~Le and Xuan Yang.
\newblock Tiny imagenet visual recognition challenge.
\newblock \emph{CS 231N}, 7\penalty0 (7):\penalty0 3, 2015.

\bibitem[Lu et~al.(2020)Lu, Nott, Olson, Todeschini, Vahabi, Carmon, and
  Schmidt]{cifar102}
Shangyun Lu, Bradley Nott, Aaron Olson, Alberto Todeschini, Hossein Vahabi,
  Yair Carmon, and Ludwig Schmidt.
\newblock Harder or different? a closer look at distribution shift in dataset
  reproduction.
\newblock In \emph{ICML Workshop on Uncertainty and Robustness in Deep
  Learning}, volume~5, pp.\ ~15, 2020.

\bibitem[McMahan \& Ramage()McMahan and Ramage]{gboard}
Brendan McMahan and Daniel Ramage.
\newblock Federated learning: Collaborative machine learning without
  centralized training data.
\newblock In \emph{Google Research Blog}.
\newblock
  \url{https://ai.googleblog.com/2017/04/federated-learning-collaborative.html}.

\bibitem[McMahan et~al.(2017)McMahan, Moore, Ramage, Hampson, and
  y~Arcas]{fedsgd}
Brendan McMahan, Eider Moore, Daniel Ramage, Seth Hampson, and
  Blaise~Ag{\"{u}}era y~Arcas.
\newblock Communication-efficient learning of deep networks from decentralized
  data.
\newblock In \emph{{AISTATS}}, 2017.

\bibitem[Melis et~al.(2019)Melis, Song, Cristofaro, and Shmatikov]{propertyInf}
Luca Melis, Congzheng Song, Emiliano~De Cristofaro, and Vitaly Shmatikov.
\newblock Exploiting unintended feature leakage in collaborative learning.
\newblock In \emph{{IEEE} Symposium on Security and Privacy}, 2019.

\bibitem[Pasquini et~al.(2022)Pasquini, Francati, and Ateniese]{pasquini}
Dario Pasquini, Danilo Francati, and Giuseppe Ateniese.
\newblock Eluding secure aggregation in federated learning via model
  inconsistency.
\newblock In \emph{{CCS}}, 2022.

\bibitem[Paulik et~al.(2021)Paulik, Seigel, Mason, Telaar, Kluivers, van Dalen,
  Lau, Carlson, Granqvist, Vandevelde, Agarwal, Freudiger, Byde, Bhowmick,
  Kapoor, Beaumont, Cahill, Hughes, Javidbakht, Dong, Rishi, and Hung]{siri}
Matthias Paulik, Matt Seigel, Henry Mason, Dominic Telaar, Joris Kluivers,
  Rogier~C. van Dalen, Chi~Wai Lau, Luke Carlson, Filip Granqvist, Chris
  Vandevelde, Sudeep Agarwal, Julien Freudiger, Andrew Byde, Abhishek Bhowmick,
  Gaurav Kapoor, Si~Beaumont, {\'{A}}ine Cahill, Dominic Hughes, Omid
  Javidbakht, Fei Dong, Rehan Rishi, and Stanley Hung.
\newblock Federated evaluation and tuning for on-device personalization: System
  design {\&} applications.
\newblock \emph{arXiv}, 2021.

\bibitem[Phong et~al.(2018)Phong, Aono, Hayashi, Wang, and
  Moriai]{analyticPhong}
Le~Trieu Phong, Yoshinori Aono, Takuya Hayashi, Lihua Wang, and Shiho Moriai.
\newblock Privacy-preserving deep learning via additively homomorphic
  encryption.
\newblock \emph{{IEEE} Trans. Inf. Forensics Secur.}, \penalty0 (5), 2018.

\bibitem[Recht et~al.(2018)Recht, Roelofs, Schmidt, and Shankar]{cifar101}
Benjamin Recht, Rebecca Roelofs, Ludwig Schmidt, and Vaishaal Shankar.
\newblock Do cifar-10 classifiers generalize to cifar-10?
\newblock \emph{arXiv preprint arXiv:1806.00451}, 2018.

\bibitem[Ronneberger et~al.(2015)Ronneberger, Fischer, and Brox]{unet}
Olaf Ronneberger, Philipp Fischer, and Thomas Brox.
\newblock U-net: Convolutional networks for biomedical image segmentation.
\newblock In \emph{Medical Image Computing and Computer-Assisted
  Intervention--MICCAI 2015: 18th International Conference, Munich, Germany,
  October 5-9, 2015, Proceedings, Part III 18}, pp.\  234--241. Springer, 2015.

\bibitem[Tschandl et~al.(2018)Tschandl, Rosendahl, and
  Kittler]{tschandl2018ham10000}
Philipp Tschandl, Cliff Rosendahl, and Harald Kittler.
\newblock The ham10000 dataset, a large collection of multi-source
  dermatoscopic images of common pigmented skin lesions.
\newblock \emph{Scientific data}, 5\penalty0 (1):\penalty0 1--9, 2018.

\bibitem[Tsipras et~al.(2018)Tsipras, Santurkar, Engstrom, Turner, and
  Madry]{resimagenet}
Dimitris Tsipras, Shibani Santurkar, Logan Engstrom, Alexander Turner, and
  Aleksander Madry.
\newblock Robustness may be at odds with accuracy.
\newblock \emph{arXiv preprint arXiv:1805.12152}, 2018.

\bibitem[Vero et~al.(2022)Vero, Balunovic, Dimitrov, and Vechev]{tableak}
Mark Vero, Mislav Balunovic, Dimitar~I. Dimitrov, and Martin~T. Vechev.
\newblock Data leakage in tabular federated learning.
\newblock \emph{ICML}, 2022.

\bibitem[Wen et~al.(2022)Wen, Geiping, Fowl, Goldblum, and Goldstein]{fishing}
Yuxin Wen, Jonas Geiping, Liam Fowl, Micah Goldblum, and Tom Goldstein.
\newblock Fishing for user data in large-batch federated learning via gradient
  magnification.
\newblock In \emph{{ICML}}, 2022.

\bibitem[Wu et~al.(2022)Wu, Zhao, Chen, and van Moorsel]{oldtabular}
Han Wu, Zilong Zhao, Lydia~Y. Chen, and Aad van Moorsel.
\newblock Federated learning for tabular data: Exploring potential risk to
  privacy.
\newblock In \emph{{ISSRE}}, 2022.

\bibitem[Wu et~al.(2021)Wu, Chen, Guo, and Weinberger]{lti}
Ruihan Wu, Xiangyu Chen, Chuan Guo, and Kilian~Q Weinberger.
\newblock Learning to invert: Simple adaptive attacks for gradient inversion in
  federated learning.
\newblock \emph{arXiv}, 2021.

\bibitem[Ye et~al.(2022)Ye, Maddi, Murakonda, Bindschaedler, and Shokri]{mia}
Jiayuan Ye, Aadyaa Maddi, Sasi~Kumar Murakonda, Vincent Bindschaedler, and Reza
  Shokri.
\newblock Enhanced membership inference attacks against machine learning
  models.
\newblock In \emph{{CCS}}, 2022.

\bibitem[Yin et~al.(2021)Yin, Mallya, Vahdat, Alvarez, Kautz, and
  Molchanov]{nvidia}
Hongxu Yin, Arun Mallya, Arash Vahdat, Jose~M. Alvarez, Jan Kautz, and Pavlo
  Molchanov.
\newblock See through gradients: Image batch recovery via gradinversion.
\newblock In \emph{{CVPR}}, 2021.

\bibitem[Yue et~al.(2022)Yue, Jin, Wong, Baron, and Dai]{study}
Kai Yue, Richeng Jin, Chau{-}Wai Wong, Dror Baron, and Huaiyu Dai.
\newblock Gradient obfuscation gives a false sense of security in federated
  learning.
\newblock \emph{arXiv}, 2022.

\bibitem[Zhang et~al.(2023)Zhang, Huang, Zhang, and Qi]{zhang23}
Shuaishuai Zhang, Jie Huang, Zeping Zhang, and Chunyang Qi.
\newblock Compromise privacy in large-batch federated learning via malicious
  model parameters.
\newblock In \emph{ICA3PP}, 2023.

\bibitem[Zhao et~al.(2020)Zhao, Mopuri, and Bilen]{iDLG}
Bo~Zhao, Konda~Reddy Mopuri, and Hakan Bilen.
\newblock idlg: Improved deep leakage from gradients.
\newblock \emph{arXiv}, 2020.

\bibitem[Zhao et~al.(2023)Zhao, Sharma, Elkordy, Ezzeldin, Avestimehr, and
  Bagchi]{mandrake}
Joshua~C. Zhao, Atul Sharma, Ahmed~Roushdy Elkordy, Yahya~H. Ezzeldin, Salman
  Avestimehr, and Saurabh Bagchi.
\newblock Secure aggregation in federated learning is not private: Leaking user
  data at large scale through model modification.
\newblock \emph{arXiv}, 2023.

\bibitem[Zhu \& Blaschko(2021)Zhu and Blaschko]{rgap}
Junyi Zhu and Matthew~B. Blaschko.
\newblock {R-GAP:} recursive gradient attack on privacy.
\newblock In \emph{{ICLR}}, 2021.

\bibitem[Zhu et~al.(2019)Zhu, Liu, and Han]{dlg}
Ligeng Zhu, Zhijian Liu, and Song Han.
\newblock Deep leakage from gradients.
\newblock In \emph{NeurIPS}, 2019.

\end{thebibliography}
\bibliographystyle{iclr2024_conference}

\message{^^JLASTREFERENCESPAGE \thepage^^J}

\ifincludeappendixx
	\clearpage
	\appendix
	\clearpage
\section{Detectability Experiments} \label{sec:app_detection}
Here we provide more details regarding our detectability experiments.

\paragraph{Measuring D-SNR}
To produce~\cref{fig:dsnr}, we consider 4 \tool{}-trained malicious models (CIFAR10, \emph{Bright}/\emph{Dark}, batch size 128/256), as well as 8 checkpoints made at various points during natural training, using the same initialization as used for \tool{}.
Then, for each value of $B \in \{16, 32, 64\}$, we choose $5$ random batches of size $B$ from the training set and $5$ random batches of size $B$ from the test set of CIFAR10.
For each batch, we compute the D-SNR on each of the 12 networks using~\cref{eq:dsnr} and plot the resulting value as a point in~\cref{fig:dsnr} (blue for natural and green for \tool{} networks).
For example disaggregation attacks, we use a publicly available implementation of the attacks of \cite{fishing} and modify the \emph{multiplier} parameter to control the strength of the attack.
We use the default setting where batches are chosen such that all images belong to the same class (\emph{car} in this case).
The three reconstructions of the example disaggregation attack are obtained by running the modernized variant of the attack of \citet{geiping} on the disaggregated batch. The modernized variant is implemented in the Breaching framework, which \citet{fishing} is a part of.
Finally, for the reconstruction of \tool{}, we aimed to show an image from the same class (a car), with D-SNR slightly below the D-SNR of the leftmost example disaggregation point ($0.72$).
To do this, we use the \emph{Dark} property and the dark car image from~\cref{fig:quality}, and sample the other $63$ examples in the batch randomly from the test set until the D-SNR falls in the $[0.62, 0.72]$ range.
We stop as soon as we find such a batch and report the reconstruction produced by~\tool{}.

\paragraph{Measuring transmission}
As noted in~\cref{sec:detection}, to be applicable to convolutional networks, boosted analytical attacks require handcrafted changes to convolutional layers that simply transmit the inputs unchanged.
While, as noted above, even in the ideal case, this cannot lead to good reconstructions, we illustrate the point that such change is detectable by defining a metric similar to D-SNR, which can be interpreted as a \emph{transmission signal-to-noise ratio}, measured on the first convolutional layer.
Namely, for each filter in the first convolutional layer, we divide the absolute values of the largest entry by the sum of the absolute values of all other entries. Intuitively, we treat the entry with the largest absolute value as the signal, and measure how well this is transmitted by the filter.
The ratio is high when the filter transmits the input unchanged, and is $\infty$ for the handcrafted changes used by the boosted analytical attacks.
We compute this metric on the 12 networks used in~\cref{fig:dsnr} (see previous paragraph) and show the results in \cref{fig:transmission}. Intuitively, the red line at $1.0$ indicates the case where there are equal amounts of the pixel being transmitted and all other pixels.
We can observe that all networks have values below $0.3$, confirming that transmission is indeed unusual and not a case that ever happens naturally, implying that if boosted analytical attacks that use this technique would be able to obtain good results, they would still be easily detectable in the weight space.

\begin{figure*}[h]  
    \centering
\includegraphics[width=0.99 \textwidth]{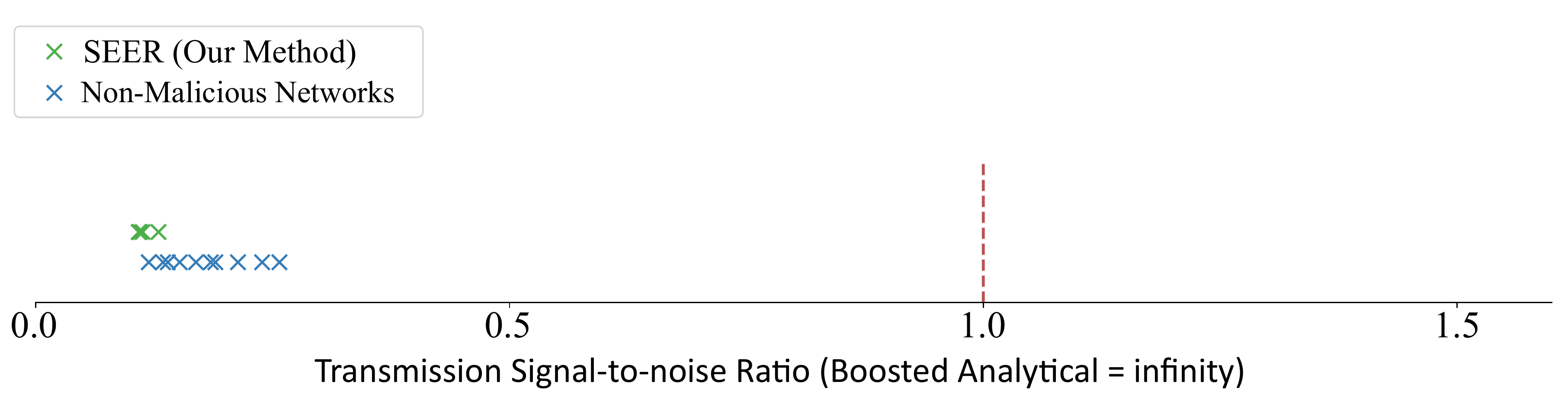}
\caption{The transmission signal-to-noise ratio of several \tool{}-trained and naturally trained networks. The same metric has a value of $\infty$ for all boosted analytical attacks.}
\vspace{-1.2em}
\label{fig:transmission}
\end{figure*}
\section{Precise Threat Model} \label{app:threat_model}
\tool{} is an MS gradient leakage attack, and thus assumes the ability of the attacker to choose the weights of the model such that client data is easier to recover. Similarly to most other MS attacks~\citep{cah,fishing,zhang23}, we focus on attacking the FedSGD~\citep{fedsgd} protocol at a single communication round, and similarly to ~\cite{fishing}, we recover a single user image from a large batch of client images. Importantly, unlike previous MS attacks, \tool{} does not lift any prior honest attack~\citep{cah,rtf,mandrake,zhang23}, does not require restrictive assumptions such as architecture tweaking~\citep{mandrake}, side-channel information, or knowledge of batch normalization data or labels~\citep{fishing}, and does not require the ability to send different updates to different users~\citep{pasquini,mandrake}. Similarly to \cite{rtf}, \tool{} requires auxiliary data to be available to the server. We believe this does not limit \tool{}'s practical applicability, as we demonstrate that we can train \tool{}'s malicious weights on a small amount of data (\cref{sec:app_size_data}) and that our attack shows remarkable transferability across datasets and data distributions (\cref{sec:experiments} and \cref{app:corruptions} and \cref{app:isic2019}). 

\section{Reconstructing from Securely-Aggregated Gradient Updates} \label{sec:app_secagg}
As described in \cref{sec:attack:components}, we have designed a more elaborate property $\mathcal{P}$ for the case of attacking securely aggregated gradient updates. Our property is based on a combination of local (in-batch) and global distribution information about the client data allowing us to handle this more complex case. In this section, we describe in detail how this is done.

As described in \cref{sec:attack:components}, we need to define the property $\mathcal{P}$ with respect to a range of brightnesses, such that $P(|I_{\text{rec}}|=1)$ is maximized. Thus, generating $\mathcal{P}$ is reduced to finding the correct threshold $\tau$ on the client image brightnesses, with which we later train \tool{}. 

To calculate the threshold $\tau$ in the multi-client setting, we use the insight that individual client batches are still generated in the presence of BN before their aggregation. To this end, we normalize the brightnesses within individual client batches of size $B$ for 20000 sampled client batches and use the sampled normalized brightness to generate the cumulative density function(CDF) of their empirical distribution. We then choose the threshold $\tau$ on this distribution to maximize the probability that exactly one out of the $C$ aggregated clients has exactly one image with normalized brightness above the threshold. 
For simplicity, we demonstrate this in the case of maximal brightness, as the minimal-brightness case is equivalent. We estimate the probability of having exactly one image with normalized brightness above the threshold as:
\begin{equation}
	(1-\Phi_1(\tau))*\Phi_2(\tau)*\Phi_1(\tau)^{C-1}\label{eq:secaggr_thresh}
\end{equation}
where $\Phi_1$ is the CDF of the top brightness in a sampled batch, and $\Phi_2$ is the CDF of the second-highest brightness in a sampled batch. The equation can be intuitively rephrased as follows---for exactly one client, the highest normalized brightness within its batch is above $\tau$, \emph{and} the second-highest brightness is below $\tau$, while for the rest, \emph{all} brightnesses are below $\tau$. To optimize \cref{eq:secaggr_thresh} for the threshold $\tau$, we use the golden section search method - a numerical optimization technique that repeatedly divides a search interval by the golden ratio to efficiently locate the (possibly-local) extremum of a function of a single variable.

\section{Existence of the Property $\mathcal{P}$}\label{sec:app_comp_ex_props}
A key assumption of our attack is the existence of an image property $\mathcal{P}$ satisfied by exactly one image $\vx\in\mathbb{R}^d$ in the client batch. We note that if we do not impose any restrictions on $\mathcal{P}$, properties of the type ``equal to $\vx$'' satisfy the requirement as they can single out any $\vx$ from any batch. However, such properties do not generalize well across batches, limiting \tool{}'s practical applicability. A more interesting question then becomes, is there always a property $\mathcal{P}$ that can separate exactly one image from a client batch that additionally is also (i) simple enough to train \tool{}'s malicious weights on well, and (ii) transferable across batches without retraining. To this end, in this section we theoretically investigate the existence of properties $\mathcal{P}$ of the type $m(\vx) < \tau$, where $m$ is a linear function and $\tau\in\mathbb{R}$ is a threshold, as in our experiments (see \cref{sec:app_props}) these types of properties show good learnability and generalizability. 

Under these assumptions, we can reformulate the question if $\mathcal{P}$ exists for a given batch and chosen image $\vx$ in it, as the question if a hyperplane $m(\vx) - \tau = 0$ exists that separates the chosen image, represented as a point in $d$-dimensional space, from the rest of the images in the batch. This question is answered positively by a well-known result from theoretical ML (see e.g., \cite{vc}) which states that the VC dimension of affine classifiers (in this case $sgn(m(\vx)-\tau)$) in $\mathbb{R}^d$ is $d+1$. Thus, a linear property $\mathcal{P}$ that is true only for the chosen images $\vx$ in the batch always exists when the batch images are in a general position (which usually holds) and $B\leq d+1$ (which holds even for low-dim images like CIFAR10, as $d+1=3073$ which is far above practical batch sizes).

\section{Possible Defenses against \tool{}}\label{app:defenses}
Here we discuss other possible defenses against \tool{}, and why we believe they are not currently effective at preventing data leakage. 
Detection based on tracking the value of the loss during training is, as noted in \cite{cah}, unlikely to work. 
This is due to the fact that in FL, per-client values can generally be noisy, even if the global loss consistently falls. 
Such detection is also made harder by the fact that \tool{} does not require application in more than one round to pose a serious threat to client privacy, and that such attacks are often applied in the first round \citep{bayesian}.
Next, \tool{} can't be easily flagged via other kinds of client-side detection, as it only modifies the weights through continuous optimization, using no obvious handcrafted patterns (as opposed to prior work analyzed in the paper). 
Finally, defenses based on differential privacy such as DP-SGD~\citep{dpsgd} require too much noise to be practical, as we also demonstrate in our experiments in \cref{app:defense:dp}. 
As we note in \cref{sec:mitigations}, we think that future principled client-side defenses can be a promising future direction. 
\subsection{Results Under Differential Privacy} \label{app:defense:dp}
\begin{table}[t]\centering	
	\caption{ Effect of gradient clipping when different maximum gradient norms $\mathcal{C}$ are applied on CIFAR10 model trained with the Red property on $B=128$. We report the percentage of well-reconstructed images~(\emph{Rec}), the average PSNR and its standard deviation across all reconstructions~(\emph{PSNR-All}), and on the top $37\%$ images~(\emph{PSNR-Top}).} \label{table:clipping}
	 
	\newcommand{\threecol}[1]{\multicolumn{2}{c}{#1}}
	\newcommand{\fivecol}[1]{\multicolumn{5}{c}{#1}}
	\newcommand{\ninecol}[1]{\multicolumn{9}{c}{#1}}
	
	\newcommand{\bsz}{Batch Size~}
	\newcommand{\certified}{{CR(\%)}}
	
	\renewcommand{\arraystretch}{1.2}
	
	\newcommand{\ccellt}[2]{\colorbox{#1}{\makebox(20,8){{#2}}}}
	\newcommand{\ccellc}[2]{\colorbox{#1}{\makebox(8,8){{#2}}}}
	\newcommand{\ccells}[2]{\colorbox{#1}{\makebox(55,8){{#2}}}}
	
	\newcommand{\temp}[1]{\textcolor{red}{#1}}
	\newcommand{\noopcite}[1]{} 
	
	\newcommand{\skiplen}{0.004\linewidth} 
	\newcommand{\rlen}{0.01\linewidth} 
	
	\resizebox{0.4\linewidth}{!}{
		\begingroup
		\setlength{\tabcolsep}{5pt} %
		\begin{tabular}{@{}lrrrrr@{}} \toprule
			
			$\mathcal{C}$ & Rec (\%) & PSNR-Top $\uparrow$ & PSNR-All $\uparrow$ \\ 
			\midrule
			1.0 & $36.2$ & $21.5\pm 1.7$ & $17.8\pm 3.4$ \\
			2.0 & $87.3$ & $25.5\pm 1.4$ & $22.3\pm 3.2$ \\
			3.0 & $92.6$ & $27.7\pm 1.1$ & $24.7\pm 3.3$ \\
			4.0 & $\bm{95.2}$ & $28.7\pm 1.0$ & $25.9\pm 3.2$ \\
			$\infty$ & $93.5$ & \bm{$31.1\pm1.2$ }& $\bm{27.8\pm4.1}$ \\
			\bottomrule
		\end{tabular}
		\endgroup
	}
\end{table}
\begin{table}[t]\centering	
	\caption{ Effect of applying DP-SGD with maximum gradient norm of $\mathcal{C}=3$ and different noise levels $\sigma$ on CIFAR10 model trained with the Red property on $B=128$. We report the percentage of well-reconstructed images~(\emph{Rec}), the average PSNR and its standard deviation across all reconstructions~(\emph{PSNR-All}), and on the top $37\%$ images~(\emph{PSNR-Top}).} \label{table:dp}
	
	\newcommand{\threecol}[1]{\multicolumn{2}{c}{#1}}
	\newcommand{\fivecol}[1]{\multicolumn{5}{c}{#1}}
	\newcommand{\ninecol}[1]{\multicolumn{9}{c}{#1}}
	
	\newcommand{\bsz}{Batch Size~}
	\newcommand{\certified}{{CR(\%)}}
	
	\renewcommand{\arraystretch}{1.2}
	
	\newcommand{\ccellt}[2]{\colorbox{#1}{\makebox(20,8){{#2}}}}
	\newcommand{\ccellc}[2]{\colorbox{#1}{\makebox(8,8){{#2}}}}
	\newcommand{\ccells}[2]{\colorbox{#1}{\makebox(55,8){{#2}}}}
	
	\newcommand{\temp}[1]{\textcolor{red}{#1}}
	\newcommand{\noopcite}[1]{} 
	
	\newcommand{\skiplen}{0.004\linewidth} 
	\newcommand{\rlen}{0.01\linewidth} 
	
	\resizebox{0.44\linewidth}{!}{
		\begingroup
		\setlength{\tabcolsep}{5pt} %
		\begin{tabular}{@{}lrrrrr@{}} \toprule
			
			$\sigma$ & Rec (\%) & PSNR-Top $\uparrow$ & PSNR-All $\uparrow$ \\ 
			\midrule
			0.0 & $\bm{92.6}$ & $\bm{27.7\pm 1.1}$ & $\bm{24.7\pm 3.3}$ \\
			\num{1e-4} & $\bm{92.6}$ & $\bm{27.7\pm 1.1}$ & $\bm{24.7\pm 3.3}$ \\
			\num{1e-3} & $92.4$ & $27.2\pm 1.0$ & $24.4\pm 3.1$ \\
			\num{3e-3} & $90.3$ & $24.6\pm 0.6$ & $22.6\pm 2.4$ \\
			\num{5e-3} & $84.0$ & $21.9\pm 0.4$ & $20.4\pm 1.9$ \\	
			\num{7e-3} & $43.5$ & $19.7\pm 0.4$ & $18.4\pm 1.8$ \\
			\num{1e-2} & $0.0$ & $17.2\pm 0.4$ & $15.3\pm 2.4$ \\
			\bottomrule
		\end{tabular}
		\endgroup
	}
\end{table}
We demonstrate \tool{}'s performance under defenses based on differential privacy. In particular, we apply our attack on gradients obtained from the DP-SGD~\citep{dpsgd} algorithm.
In DP-SGD, the clients defend their data by first clipping the norms of the per-layer gradients of each of their data points to at most $\mathcal{C}$, and then adding Gaussian noise with standard deviation of $\mathcal{C}\cdot\sigma$ to them. In order to better understand what effect those two components of the defense have on our method, in \cref{table:clipping} we experiment with different clipping norms $\mathcal{C}$ for $\sigma=0$ (\ie no noise added), while in \cref{table:dp} we experiment with the noise strength $\sigma$ for clipping norm of $3$. We use $\mathcal{C}=3$ as \cite{dpsgd} recommends that value for CIFAR10, the dataset we experiment with. Both experiments were conducted on a \tool{} model trained with $B=128$ and the Red property.

In our experiments, the performance of \tool{} increased when we explicitly took into account the use of DP-SGD during our attack procedure. In particular, before applying the attack in all experiments we first estimate the median clipping factor for each layer individually on $1000$ batches of size $128$ taken from our auxiliary data. These approximate factors are then reapplied to the total gradients sent from the clients to the malicious server before applying \cref{alg:toolattack}. 

The results in \cref{table:clipping} and \cref{table:dp} confirm the trends observed in prior work \citep{dlg,geiping,bayesian} that low clipping norms $\mathcal{C}$ and high noise levels $\sigma$ result in more effective defense mechanisms. Still, we observe that \tool{} is fairly robust to DP-SGD, as even for clipping norm of $2$, and high noise levels of $\num{5e-3}$ our method is able to recover private date from more than $85\%$ of client batches.

\section{Additional Experiments}\label{sec:app_exp}
In this section, we provide additional experimental results, which we did not include in the main body due to space constraints.
\subsection{Extended Large Batch Experiments}\label{sec:app_full_exp}
\begin{table*}[t]\centering
	
	\caption{Large batch reconstruction on bright and red properties from batches of different sizes $B$ on CIFAR10. We report the percentage of well-reconstructed images~(\emph{Rec}), the average PSNR and its standard deviation on all reconstructions~(\emph{PSNR-All}), and across the top $37\%$ images~(\emph{PSNR-Top}).} \label{table:single_c10_full}
	
	\newcommand{\threecol}[1]{\multicolumn{3}{c}{#1}}
	\newcommand{\fivecol}[1]{\multicolumn{5}{c}{#1}}
	\newcommand{\ninecol}[1]{\multicolumn{9}{c}{#1}}
	
	\newcommand{\bsz}{Batch Size~}
	\newcommand{\certified}{{CR(\%)}}
	
	\renewcommand{\arraystretch}{1.2}
	
	\newcommand{\ccellt}[2]{\colorbox{#1}{\makebox(20,8){{#2}}}}
	\newcommand{\ccellc}[2]{\colorbox{#1}{\makebox(8,8){{#2}}}}
	\newcommand{\ccells}[2]{\colorbox{#1}{\makebox(55,8){{#2}}}}
	
	\newcommand{\temp}[1]{\textcolor{red}{#1}}
	\newcommand{\noopcite}[1]{} 
	
	\newcommand{\skiplen}{-0.4em} 
	\newcommand{\rlen}{0.01\linewidth} 
	
	\resizebox{0.8 \linewidth}{!}{
		\begingroup
		\setlength{\tabcolsep}{5pt}
		\begin{tabular}{@{}l rrr p{\skiplen}  rrr@{}} \toprule
			
			& \threecol{CIFAR10, Red} && \threecol{CIFAR10, Bright}\\
			
			\cmidrule(l{5pt}r{5pt}){2-4} \cmidrule(l{5pt}r{5pt}){6-8}
			
			$B$&  Rec (\%) & PSNR-Top $\uparrow$ & PSNR-All $\uparrow$ &&  Rec (\%) & PSNR-Top $\uparrow$ & PSNR-All $\uparrow$ \\ \midrule
			64 & $87.3$ & $30.4\pm1.1$ & $26.5\pm5.2$ && $89.4$ & $32.1\pm2.0$ & $27.2\pm5.3$ \\
			128 & $93.5$ & $31.1\pm1.2$ & $27.8\pm4.1$ && $\bm{94.2}$ & $31.9\pm1.7$ & $28.2\pm4.3$ \\
			256 & \bm{$94.7$} & \bm{$31.3\pm1.0$} & \bm{$28.0\pm4.0$} && $93.5$ & $\bm{32.8\pm2.0}$ & $\bm{28.5\pm5.0}$ \\
			512 & $94.4$ & $30.0\pm1.2$ & $26.6\pm3.8$ && $87.8$ & $26.6\pm1.8$ & $23.2\pm3.5$\\
			\bottomrule
		\end{tabular}
		\endgroup
	} 
\end{table*}
\begin{table*}[t]\centering
	
	\caption{Large batch reconstruction on bright and red properties from batches of different sizes $B$ on CIFAR100. We report the percentage of well-reconstructed images~(\emph{Rec}), the average PSNR and its standard deviation across all reconstructions~(\emph{PSNR-All}), and on the top $37\%$ images~(\emph{PSNR-Top}).} \label{table:single_c100_full}
	
	\newcommand{\threecol}[1]{\multicolumn{3}{c}{#1}}
	\newcommand{\fivecol}[1]{\multicolumn{5}{c}{#1}}
	\newcommand{\ninecol}[1]{\multicolumn{9}{c}{#1}}
	
	\newcommand{\bsz}{Batch Size~}
	\newcommand{\certified}{{CR(\%)}}
	
	\renewcommand{\arraystretch}{1.2}
	
	\newcommand{\ccellt}[2]{\colorbox{#1}{\makebox(20,8){{#2}}}}
	\newcommand{\ccellc}[2]{\colorbox{#1}{\makebox(8,8){{#2}}}}
	\newcommand{\ccells}[2]{\colorbox{#1}{\makebox(55,8){{#2}}}}
	
	\newcommand{\temp}[1]{\textcolor{red}{#1}}
	\newcommand{\noopcite}[1]{} 
	
	\newcommand{\skiplen}{-0.4em} 
	\newcommand{\rlen}{0.01\linewidth} 
	
	\resizebox{0.8 \linewidth}{!}{
		\begingroup
		\setlength{\tabcolsep}{5pt}
		\begin{tabular}{@{}l rrr p{\skiplen}  rrr@{}} \toprule
			
			& \threecol{CIFAR100, Red} && \threecol{CIFAR100, Bright}\\
			
			\cmidrule(l{5pt}r{5pt}){2-4} \cmidrule(l{5pt}r{5pt}){6-8}
			
			$B$&  Rec (\%) & PSNR-Top $\uparrow$ & PSNR-All $\uparrow$ &&  Rec (\%) & PSNR-Top $\uparrow$ & PSNR-All $\uparrow$ \\ \midrule
			64 & $97.1$ & $31.7\pm1.1$ & $29.0\pm3.4$ && $95.6$ & $32.2\pm1.5$ & $28.2\pm4.3$\\
			128 & $97.4$ & $31.8\pm1.1$ & $29.3\pm3.2$ && $94.7$ & $30.0\pm1.3$ & $26.5\pm3.7$\\
			256 & $97.7$ & $31.3\pm1.0$ & $28.6\pm3.2$ && $\bm{98.1}$ & $\bm{35.2\pm1.4}$ & $\bm{30.8\pm4.8}$\\
			512& $\bm{98.6}$ & $\bm{33.1\pm1.1}$ & $\bm{30.5\pm3.1}$   && $95.0$ & $32.2\pm1.6$ & $27.6\pm4.6$\\
			\bottomrule
		\end{tabular}
		\endgroup
	} 
\end{table*}
In \cref{table:single_c10_full} and \cref{table:single_c100_full}, we present the extended version of our CIFAR10 and CIFAR100 single-client experiments, first presented in \cref{sec:experiments}. We observe similar trends as in the original experiments. For example, we observe that CIFAR10 and CIFAR100 performances are similar and that there is no major difference in performance between the most bright and most red image properties. 
\subsection{Extended Secure Aggregation Experiments}\label{sec:app_full_aggr_exp}
\begin{table*}[t]\centering

	\caption{Reconstructions on securely aggregated batches on the bright and dark properties with different numbers of clients $C$ on CIFAR10, for different total numbers of images. We report the percentage of correctly reconstructed images (\emph{Rec}) and the average PSNR across the top $37\%$ images (\emph{PSNR-Top}).} \label{table:full_secure}
	
	\newcommand{\threecol}[1]{\multicolumn{2}{c}{#1}}
	\newcommand{\fivecol}[1]{\multicolumn{5}{c}{#1}}
	\newcommand{\ninecol}[1]{\multicolumn{9}{c}{#1}}
	
	\newcommand{\bsz}{Batch Size~}
	\newcommand{\certified}{{CR(\%)}}
	
	\renewcommand{\arraystretch}{1.2}
	
	\newcommand{\ccellt}[2]{\colorbox{#1}{\makebox(20,8){{#2}}}}
	\newcommand{\ccellc}[2]{\colorbox{#1}{\makebox(8,8){{#2}}}}
	\newcommand{\ccells}[2]{\colorbox{#1}{\makebox(55,8){{#2}}}}
	
	\newcommand{\temp}[1]{\textcolor{red}{#1}}
	\newcommand{\noopcite}[1]{} 
	
	\newcommand{\skiplen}{-0.1em} 
	\newcommand{\rlen}{0.01\linewidth} 
	
	\resizebox{1.0 \linewidth}{!}{
		\begingroup
		\setlength{\tabcolsep}{5pt} %
		\begin{tabular}{@{}l rr p{\skiplen}  rr p{\skiplen} rr p{\skiplen}  rr p{\skiplen}  rr  p{\skiplen}  rr@{}} \toprule

			& \threecol{$C=4$, Dark} && \threecol{$C=4$, Bright} && \threecol{$C=8$, Dark} && \threecol{$C=8$, Bright}\\
			
			\cmidrule(l{5pt}r{5pt}){2-3} \cmidrule(l{5pt}r{5pt}){5-6} \cmidrule(l{5pt}r{5pt}){8-9} \cmidrule(l{5pt}r{1pt}){11-12}
			
			\#Imgs&  Rec (\%) & PSNR-Top&&  Rec (\%) & PSNR-Top&& Rec (\%) & PSNR-Top && Rec (\%) & PSNR-Top  \\ \midrule
			64 & $50.2$ & $27.5\pm3.0$ && $41.4$ & $\bm{27.3\pm3.1}$ && $43.0$ & $27.4\pm3.3$ && $41.3$ & $26.6\pm3.7$ \\
			128 & $51.3$ & $28.8\pm2.6$ && $44.2$ & $26.8\pm3.0$ && $43.4$ & $27.6\pm3.5$ && $40.6$ & $\bm{27.3\pm3.3}$ \\
			256 & $50.9$ & $29.8\pm2.3$ && $51.9$ & $\bm{27.3\pm2.5}$ && $51.7$ & $27.0\pm2.9$ && $41.9$ & $25.4\pm3.1$ \\
			512 & $\bm{61.3}$ & $\bm{30.2\pm2.4}$ && $\bm{52.9}$ & $25.7\pm2.4$ && $\bm{56.3}$ & $\bm{28.7\pm2.9}$ && $\bm{51.7}$ & $25.9\pm2.8$ \\

			\bottomrule
		\end{tabular}
		\endgroup
	}
\end{table*}
In \cref{table:full_secure}, we present an extended version of our CIFAR10 multi-client experiments first presented in \cref{sec:experiments} containing results for both the bright and dark properties. While we see the dark property reconstructions are generally slightly better than the bright ones, we observe similar trends between the two sets of the experiments. This suggests our method for attacking secure aggregation federated updates is effective regardless the property used.
\subsection{Robustness to $B$}\label{sec:app_robustness}
\begin{table*}[t]\centering
	\caption{Large batch reconstruction on the bright property on CIFAR10 on a network trained with batch size $B=128$ and tested for various client batch sizes $B_\text{test}$. We report the percentage of well-reconstructed images~(\emph{Rec}), the average PSNR and its standard deviation on all reconstructions~(\emph{PSNR-All}), and across the top $37\%$ images~(\emph{PSNR-Top}).} \label{table:robustness}
	
	\newcommand{\threecol}[1]{\multicolumn{2}{c}{#1}}
	\newcommand{\fivecol}[1]{\multicolumn{5}{c}{#1}}
	\newcommand{\ninecol}[1]{\multicolumn{9}{c}{#1}}
	
	\newcommand{\bsz}{Batch Size~}
	\newcommand{\certified}{{CR(\%)}}
	
	\renewcommand{\arraystretch}{1.2}
	
	\newcommand{\ccellt}[2]{\colorbox{#1}{\makebox(20,8){{#2}}}}
	\newcommand{\ccellc}[2]{\colorbox{#1}{\makebox(8,8){{#2}}}}
	\newcommand{\ccells}[2]{\colorbox{#1}{\makebox(55,8){{#2}}}}
	
	\newcommand{\temp}[1]{\textcolor{red}{#1}}
	\newcommand{\noopcite}[1]{} 
	
	\newcommand{\skiplen}{0.004\linewidth} 
	\newcommand{\rlen}{0.01\linewidth} 
	
	\resizebox{0.5 \linewidth}{!}{
		\begingroup
		\setlength{\tabcolsep}{5pt} %
		\begin{tabular}{@{}l rrr@{}} \toprule
			
			$B_\text{test}$ & Rec (\%) & PSNR-Top $\uparrow$ & PSNR-All $\uparrow$\\ \midrule
			64 & 42.0\% & $21.51\pm1.83$ & $18.51\pm2.93$ \\
			96 & 87.4\% & $30.56\pm1.31$ & $26.28\pm4.92$ \\
			128 & \textbf{94.2\%} & \bm{$31.91\pm1.73$} & \bm{$28.15\pm4.34$} \\	
			192 & 86.3\% & $30.78\pm2.37$ & $25.77\pm5.14$ \\	
			256 & 67.5\% & $28.02\pm2.79$ & $22.42\pm5.14$ \\
			\bottomrule
		\end{tabular}
		\endgroup
	}
\end{table*}
In this section, we demonstrate that attack parameters $\theta_f$ generated by \tool{} for a particular client batch size $B$ can work to a large extent for batch sizes close to the original one, thus relaxing the requirement that the exact client batch size $B$ is known during the crafting of the malicious model $f$. In particular, in~\cref{table:robustness}, we show the effect of applying our single-client attack trained on $B = 128$ on CIFAR10 using the \emph{Bright} image property for clients with varying batch sizes $B_\text{test}$. We observe that while, as expected, \tool{} performs best when $B_\text{test} = B$, both the success rate and the quality of reconstruction on clients with batch sizes even $2\times$ larger than the trained one remain very good. We note that \cref{table:robustness} suggests that underestimating the client batch size $B_\text{test}$ during the training of $f$ is better than overestimating it, as the reconstruction performance when $B_\text{test} = 64$ is significantly worse than when $B_\text{test} = 256$. This is mostly caused by $d$ filtering out all images in the client batches resulting in the removal of all the client data. 
\subsection{Robustness to Image Corruptions}\label{app:corruptions}
\begin{table*}[t]\centering
	\caption{Large reconstruction in the presence of different common corruptions applied on CIFAR10 network trained on $B=128$ using the red property. As a baseline, our attack achieves Rec: $93.5\%$, PSNR-Top: $31.1\pm1.2$, and PSNR-All: $27.8\pm4.1$ .  We report the percentage of well-reconstructed images~(\emph{Rec}), the average PSNR and its standard deviation across all reconstructions~(\emph{PSNR-All}), and on the top $37\%$ images~(\emph{PSNR-Top}) for three different degrees of severity of the corruption~(\emph{Severity}).} \label{table:common_corr}
        \newcommand{\threecol}[1]{\multicolumn{3}{c}{#1}}
        \newcommand{\fivecol}[1]{\multicolumn{5}{c}{#1}}
        \newcommand{\ninecol}[1]{\multicolumn{9}{c}{#1}}
         
        \newcommand{\bsz}{Batch Size~}
        \newcommand{\certified}{{CR(\%)}}
        
        \renewcommand{\arraystretch}{1.2}
        
        \newcommand{\ccellt}[2]{\colorbox{#1}{\makebox(20,8){{#2}}}}
        \newcommand{\ccellc}[2]{\colorbox{#1}{\makebox(8,8){{#2}}}}
        \newcommand{\ccells}[2]{\colorbox{#1}{\makebox(55,8){{#2}}}}
        
        \newcommand{\temp}[1]{\textcolor{red}{#1}}
        \newcommand{\noopcite}[1]{} 
         
        \newcommand{\skiplen}{-0.2\tabcolsep} 
        \newcommand{\rlen}{0.01\linewidth} 
         
        \resizebox{\linewidth}{!}{
            \begingroup
            \setlength{\tabcolsep}{5pt} %
            \begin{tabular}{@{}l@{\hspace{1.8\tabcolsep}}rrrr rrrr rrrr@{}} \toprule

                & \threecol{Severity $=1$} && \threecol{Severity $=3$} && \threecol{Severity $=5$} \\
                
                \cmidrule(l{5pt}r{5pt}){2-4} \cmidrule(l{5pt}r{5pt}){6-8} \cmidrule(l{5pt}r{5pt}){10-12}
                
                Corruption &  Rec (\%) & PSNR-Top $\uparrow$ & PSNR-All $\uparrow$ &&  Rec (\%) & PSNR-Top $\uparrow$ & PSNR-All $\uparrow$ && Rec (\%) & PSNR-Top $\uparrow$ & PSNR-All $\uparrow$\\ \midrule
				Brightness & $94.7$ & $\bm{31.4 \pm 1.1}$ & $\bm{28.0 \pm 4.1}$ && $92.3$ & $31.6 \pm 1.1$ & $27.7 \pm 4.5$ && $94.0$ & $\bm{31.0 \pm 1.4}$ & $27.3 \pm 4.3$\\
				Contrast & $\bm{96.2}$ & $24.6 \pm 1.0$ & $22.9 \pm 2.0$ && $2.8$ & $17.8 \pm 0.9$ & $16.6 \pm 1.2$ && $0.6$ & $15.1 \pm 1.3$ & $13.6 \pm 1.6$\\
				Contrast Fixed & $\bm{97.5}$ & $34.4 \pm 1.3$ & $30.6 \pm 4.5$ && $99.0$ & $32.2 \pm 1.5$ & $29.2 \pm 3.29$ && $98.0$ & $28.2 \pm 1.6$ & $25.4 \pm 2.9$\\
				Defocus Blur & $94.2$ & $30.2 \pm 1.1$ & $27.3 \pm 3.7$ && $94.9$ & $27.5 \pm 1.1$ & $25.1 \pm 2.9$ && $94.8$ & $25.2 \pm 0.9$ & $23.2 \pm 2.3$\\
				Elastic Transform & $95.0$ & $29.0 \pm 1.0$ & $26.3 \pm 3.3$ && $94.7$ & $27.7 \pm 1.1$ & $25.2 \pm 2.9$ && $95.3$ & $28.1 \pm 1.1$ & $25.5 \pm 3.0$\\
				Fog & $95.3$ & $26.1 \pm 1.1$ & $24.0 \pm 2.3$ && $68.6$ & $21.4 \pm 1.1$ & $19.8 \pm 1.6$ && $23.0$ & $19.5 \pm 0.9$ & $18.1 \pm 1.3$\\
				Fog Fixed & $95.4$ & $33.6 \pm 1.4$ & $29.4 \pm 4.7$ && $98.0$ & $34.6 \pm 1.2$ & $31.0 \pm 4.4$ && $97.6$ & $35.0 \pm 1.0$ & $31.1 \pm 4.7$\\
				Frost & $94.4$ & $30.2 \pm 0.9$ & $27.0 \pm 3.8$ && $94.6$ & $28.1 \pm 1.0$ & $25.4 \pm 3.3$ && $94.2$ & $26.5 \pm 0.7$ & $24.3 \pm 2.6$\\
				Gaussian Blur & $94.2$ & $30.3 \pm 1.1$ & $27.3 \pm 3.7$ && $95.0$ & $26.5 \pm 1.0$ & $24.3 \pm 2.6$ && $95.0$ & $24.3 \pm 1.0$ & $22.5 \pm 2.1$\\
				Gaussian Noise & $94.2$ & $30.3 \pm 0.8$ & $27.2 \pm 3.9$ && $92.8$ & $27.8 \pm 0.5$ & $25.2 \pm 3.4$ && $90.9$ & $26.3 \pm 0.4$ & $24.0 \pm 3.1$\\
				Glass Blur & $93.3$ & $31.2 \pm 1.3$ & $27.7 \pm 4.2$ && $95.0$ & $29.6 \pm 1.1$ & $26.8 \pm 3.5$ && $94.9$ & $29.8 \pm 1.1$ & $27.0 \pm 3.4$\\
				Impulse Noise & $93.0$ & $29.5 \pm 0.8$ & $26.6 \pm 3.7$ && $90.9$ & $26.4 \pm 0.7$ & $24.0 \pm 3.0$ && $85.8$ & $22.5 \pm 0.5$ & $20.9 \pm 2.1$\\
				Jpeg Compression & $94.6$ & $30.9 \pm 1.1$ & $27.7 \pm 3.9$ && $\bm{95.3}$ & $30.8 \pm 1.1$ & $27.7 \pm 3.8$ && $94.5$ & $30.6 \pm 1.1$ & $\bm{27.6 \pm 3.8}$\\
				Motion Blur & $95.3$ & $28.6 \pm 1.0$ & $26.1 \pm 3.1$ && $95.0$ & $26.2 \pm 1.1$ & $24.1 \pm 2.5$ && $94.6$ & $25.3 \pm 1.2$ & $23.3 \pm 2.3$\\
				Pixelate & $93.3$ & $30.8 \pm 1.1$ & $27.6 \pm 3.9$ && $94.2$ & $30.4 \pm 1.1$ & $27.3 \pm 3.8$ && $94.4$ & $29.1 \pm 1.0$ & $26.5 \pm 3.4$\\
				Saturate & $93.5$ & $29.1 \pm 1.2$ & $26.1 \pm 3.6$ && $95.0$ & $\bm{31.7 \pm 1.5}$ & $\bm{28.0 \pm 4.2}$ && $91.4$ & $24.7 \pm 0.8$ & $22.7 \pm 2.3$\\
				Shot Noise & $94.2$ & $30.7 \pm 0.9$ & $27.6 \pm 3.9$ && $92.9$ & $29.0 \pm 0.7$ & $26.2 \pm 3.6$ && $91.5$ & $27.0 \pm 0.6$ & $24.6 \pm 3.3$\\
				Snow & $93.3$ & $31.3 \pm 1.1$ & $27.9 \pm 4.2$ && $92.6$ & $30.8 \pm 0.9$ & $27.3 \pm 4.2$ && $91.8$ & $29.3 \pm 1.1$ & $25.9 \pm 4.0$\\
				Spatter & $93.3$ & $31.3 \pm 1.1$ & $27.9 \pm 4.2$ && $92.5$ & $31.1 \pm 1.0$ & $27.5 \pm 4.4$ && $93.0$ & $30.6 \pm 1.1$ & $27.1 \pm 4.2$\\
				Speckle Noise & $93.8$ & $30.8 \pm 1.0$ & $27.6 \pm 3.9$ && $92.5$ & $29.5 \pm 0.8$ & $26.5 \pm 3.9$ && $91.2$ & $27.0 \pm 0.8$ & $24.5 \pm 3.3$\\
				Zoom Blur & $93.6$ & $27.9 \pm 1.1$ & $25.3 \pm 3.2$ && $93.6$ & $26.6 \pm 1.0$ & $24.3 \pm 2.8$ && $\bm{95.4}$ & $25.5 \pm 1.0$ & $23.5 \pm 2.3$\\
                \bottomrule
            \end{tabular}
            \endgroup
        }
    \end{table*}

In this section, we show that \tool{} is robust to distributional shifts caused by common image corruptions.
To this end, we use a \tool{} model trained on the CIFAR10 trainset with the red property, batch size $B = 128$, and without secure aggregation to attack batches sampled from the CIFAR10-C dataset~\citep{common_corruptions}, where 19 different image corruptions are applied at different levels of severity (1-5) to the original testset of CIFAR10. We show the results in \cref{table:common_corr}.

We see that for all image corruptions but \emph{Fog} and \emph{Contrast}, even at severity 5, we recover images from more than 85\% of client batches with good quality (average PSNR > 23 in most cases). For \emph{Fog} and, to even greater extend, \emph{Contrast}, however, we observed that reconstructions, while preserving the image semantic, became too bright resulting in a very low PSNR numbers. To this end, we created a modified version of our attack that approximates and applies a multiplicative factor $\beta$ by which one needs to multiply the recovered normalized images such that 90\% of the recovered images after denormalization are inside the range $[0,1]$. We use 90\% to account for the fact that not all images will be correctly recovered, and we don't want these to affect our $\beta$ estimations. The results are shown under \emph{Fog Fixed} and \emph{Contrast Fixed}, where we get even better results than on the original data. We conjecture this is due to the fact that most of the corrupted images have narrower range of pixel values making obtaining high PSNR numbers easier. All in all, our experiments show that \tool{} is very robust to image corruptions, and that even the most severe changes like the ones in CIFAR10-C, sometimes resulting in hard to recognize images, can be handled well by our algorithm.
\subsection{Comparison between Our Large Batch and Secure Aggregation Properties}\label{sec:prob_vs_rel}
In this section, we compare our two \tool{} variants---one based purely on the local distribution of $m$ for the large batch setting, referred to as local $\mathcal{P}$, and the other for secure aggregation setting, described in \cref{sec:app_secagg}, which is based on a mix of the global and local distributions of $m$ and is referred to as global $\mathcal{P}$. We mount both variants of \tool{} on gradients coming from a single client with batch size $B = 128$ on CIFAR10. We note that both methods are well-defined in this setting, and either one can successfully reconstruct data from the client batches.

The results are depicted in \cref{table:compare}. While both methods successfully reconstruct the majority of client batches, we clearly see the benefits of using the local property $\mathcal{P}$. In particular, the results in \cref{table:compare} suggest that the local distribution approach reconstructs up to $1.75$ times more images, while also producing higher PSNR values not only on the full set of reconstructed batches but also on the top $37\%$ of them. This motivates the need for our single-client attack variant, and demonstrates that secure aggregation provides additional protection to individual clients.

\begin{table}\centering
	
	\caption{Large batch reconstruction on the bright and dark properties from batches of size $B = 128$ on CIFAR10 using local (\emph{Local}) and global properties $\mathcal{P}$ (\emph{Global}). We report the percentage of well-reconstructed images~(\emph{Rec}), the average PSNR and its standard deviation across all reconstructions~(\emph{PSNR-All}), and across the top $37\%$ images~(\emph{PSNR-Top}). } \label{table:compare}
	
	\newcommand{\threecol}[1]{\multicolumn{3}{c}{#1}}
	\newcommand{\fivecol}[1]{\multicolumn{5}{c}{#1}}
	\newcommand{\ninecol}[1]{\multicolumn{9}{c}{#1}}
	
	\newcommand{\bsz}{Batch Size~}
	\newcommand{\certified}{{CR(\%)}}
	
	\renewcommand{\arraystretch}{1.2}
	
	\newcommand{\ccellt}[2]{\colorbox{#1}{\makebox(20,8){{#2}}}}
	\newcommand{\ccellc}[2]{\colorbox{#1}{\makebox(8,8){{#2}}}}
	\newcommand{\ccells}[2]{\colorbox{#1}{\makebox(55,8){{#2}}}}
	
	\newcommand{\temp}[1]{\textcolor{red}{#1}}
	\newcommand{\noopcite}[1]{} 
	
	\newcommand{\skiplen}{-0.4em} 
	\newcommand{\rlen}{0.01\linewidth} 
	
	\resizebox{0.8 \linewidth}{!}{ 
		\begingroup
		\setlength{\tabcolsep}{5pt}
		\begin{tabular}{@{}l rrr p{\skiplen}  rrr@{}} \toprule

			& \threecol{CIFAR10, Bright} && \threecol{CIFAR10, Dark}\\
			
			\cmidrule(l{5pt}r{5pt}){2-4} \cmidrule(l{5pt}r{5pt}){6-8}
			
			$\mathcal{P}$ &  Rec (\%) & PSNR-Top $\uparrow$ & PSNR-All $\uparrow$ &&   Rec (\%) & PSNR-Top $\uparrow$ & PSNR-All $\uparrow$ \\ \midrule
			Global & $54.4$ & $27.0\pm1.8$ & $20.6\pm5.8$ && $61.9$ & $27.7\pm2.2$ & $21.1\pm6.1$\\
			Local & $\bm{94.2}$ & $\bm{31.9\pm1.7}$ & $\bm{28.2\pm4.3}$ && $\bm{81.3}$ & $\bm{33.6\pm1.4}$ & $\bm{27.4\pm7.3}$\\
			
			\bottomrule
		\end{tabular}
		\endgroup
	} 
\end{table}
\subsection{Additional Types of Property Metrics $m$}\label{sec:app_props}
\begin{table}\centering
	\caption{
	Large batch reconstruction on CIFAR10 with $B=128$ with different properties $\mathcal{P}$. We report the percentage of well-reconstructed images~(\emph{Rec}), the average PSNR and its standard deviation on all reconstructions~(\emph{PSNR-All}), and across the top $37\%$ images~(\emph{PSNR-Top}).} \label{table:properties}
	
	\newcommand{\threecol}[1]{\multicolumn{2}{c}{#1}}
	\newcommand{\fivecol}[1]{\multicolumn{5}{c}{#1}}
	\newcommand{\ninecol}[1]{\multicolumn{9}{c}{#1}}
	
	\newcommand{\bsz}{Batch Size~}
	\newcommand{\certified}{{CR(\%)}}
	
	\renewcommand{\arraystretch}{1.2}
	
	\newcommand{\ccellt}[2]{\colorbox{#1}{\makebox(20,8){{#2}}}}
	\newcommand{\ccellc}[2]{\colorbox{#1}{\makebox(8,8){{#2}}}}
	\newcommand{\ccells}[2]{\colorbox{#1}{\makebox(55,8){{#2}}}}
	
	\newcommand{\temp}[1]{\textcolor{red}{#1}}
	\newcommand{\noopcite}[1]{} 
	
	\newcommand{\skiplen}{0.004\linewidth} 
	\newcommand{\rlen}{0.01\linewidth} 
	
	\resizebox{0.5\linewidth}{!}{
		\begingroup
		\setlength{\tabcolsep}{5pt} %
		\begin{tabular}{@{}l rrrr@{}} \toprule
			
			Property & Rec (\%) & PSNR-Top $\uparrow$ & PSNR-All $\uparrow$\\ \midrule
			Bright & $94.2$ & $31.9\pm1.7$ & $28.2\pm4.3$ \\
			Dark & $81.3$ & $\bm{33.6\pm1.4}$ & $27.4\pm7.3$ \\ %
			Red & $93.5$ & $31.1\pm1.2$ & $27.8\pm4.1$ \\ %
			Blue & $97.2$ & $31.5\pm0.9$ & $28.6\pm3.5$ \\ %
			Green & $96.7$ & $32.8\pm1.1$ & $\bm{29.6\pm4.0}$ \\ %
			H Edge & $80.1$ & $29.0\pm1.1$ & $24.4\pm5.5$ \\ %
			V Edge & $85.8$ & $29.6\pm1.0$ & $25.5\pm5.0$ \\ %
			Green V Edge & $95.1$ & $32.5\pm1.1$ & $28.6\pm4.5$ \\ %
			Rand & $\bm{97.5}$ & $32.8\pm1.1$ & $29.4\pm3.8$ \\ %
			\bottomrule
		\end{tabular}
		\endgroup
	}
\end{table}

\begin{figure}[!ht]
	\centering
	\includegraphics[width=0.95\linewidth]{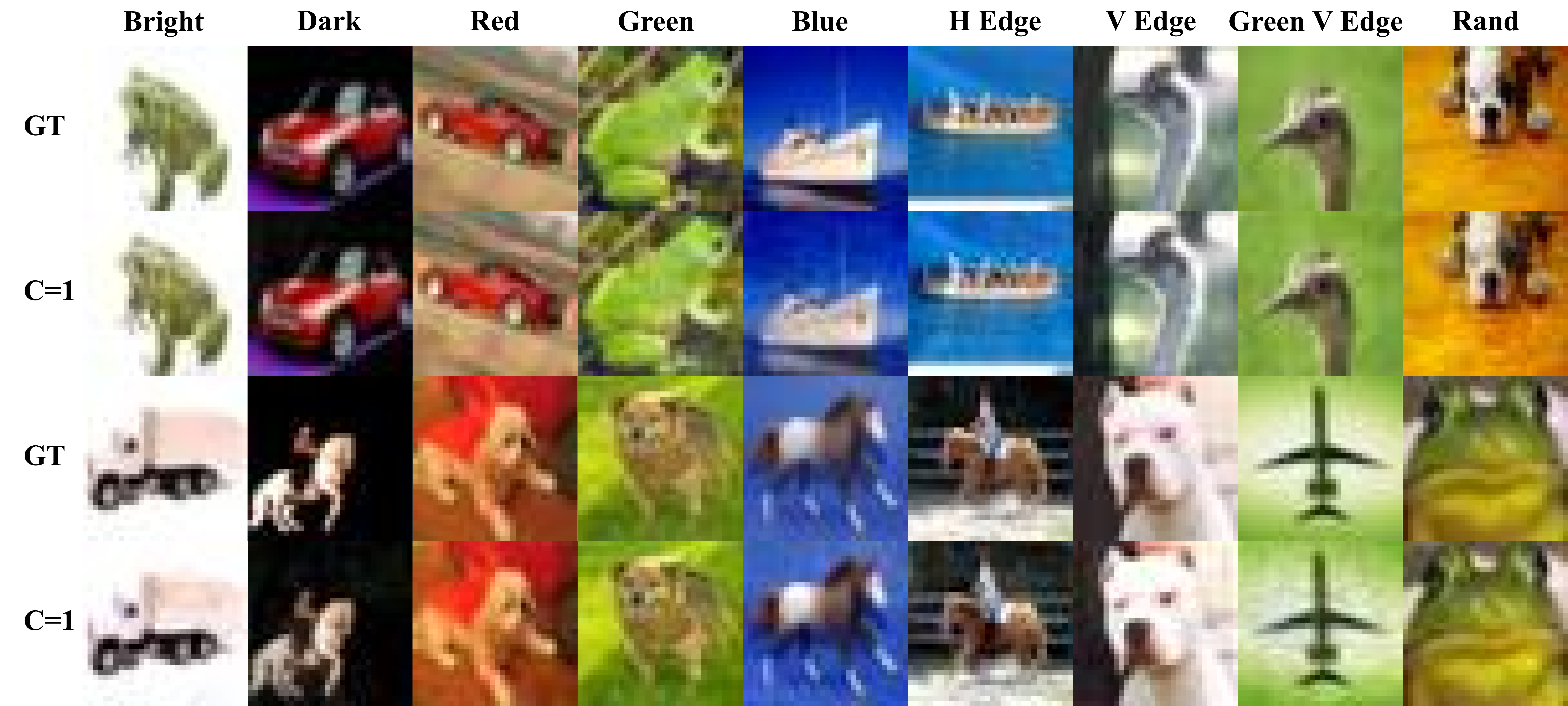}
	\caption{Example reconstructions of \tool{} trained on CIFAR10 with $C=1, B=128$ and different properties.}
	\label{fig:properties}
\end{figure}    

In this section, we demonstrate that our method works well for variety of properties $\mathcal{P}$ based on different metrics $m$.
In particular, we look at local properties $\mathcal{P}$ based on: (i) the image brightness---the most bright (\emph{Bright}) and the most dark (\emph{Dark}) image in a batch; (ii) the image color---the most red (\emph{Red}), blue (\emph{Blue}), or green (\emph{Green}) image in a batch; (iii) edges in the image---the image with the strongest horizontal (\emph{H Edge}), or vertical (\emph{V Edge}) edges in a batch; (iv) combination of image color and edges---the most green image with vertical edges (\emph{Green V Edge}); and, finally, (iv) based on random property (\emph{Rand}). For the color properties we rank the batch images based on the difference between two times the average color channel response for the chosen color and the sum of the other two average color channel responses. For the edge properties, we ranked the batch images based on the average response with the $[1,-1]$ edge filter (either in horizontal or vertical direction) on a grayscale version of the image. Further, for the combination filter we ranked images based on sum of the color and edge property scores. Finally, for the random property we ranked the batch images based on the average response to a random $3\times 3$ convolution filter that was normalized. The results for CIFAR10 for the large batch size setting for $B=128$ is shown in \cref{table:properties}. Further, example reconstructions are given in \cref{fig:properties}. 

We observe that for all properties \tool{} successfully recovers data from a large portion of the client batches (>80\%), with good quality (PSNR>24). Yet, we still observe some variability across the properties with the Random property being the easiest to attack, as demonstrated by the percentage of recovered images and the very good PSNR metrics. This is in line with the observations in \cite{rtf}. We also observed that the Dark property produces the best image reconstructions, as shown by PSNR-Top, but fails to reconstruct as often. In practice, we observed this happens due to \tool{} recovering completely black images. We conjecture this is due to lack of diversity in the images \tool{} sees during training as most images in CIFAR10 with the Dark property have completely black background. Finally, we observe that color properties are easier to attack compared to edge ones but, interestingly, when combined, like in \emph{Green V Edge}, the results become much closer to the color version. We conjecture this is due images with very pronounced colors being more similar to each other and, thus, easier to distinguish from the rest of the images in the batch.
\subsection{Results Under Trained Encoder}
\begin{table}[t]\centering	
	\caption{Large batch reconstruction on CIFAR10 model trained with the Bright property on $B=128$ after several rounds of federated training (\emph{\#Rounds}) with different learning rates (\emph{Learning Rate}) using 8 clients. We report the percentage of well-reconstructed images~(\emph{Rec}), the average PSNR and its standard deviation across all reconstructions~(\emph{PSNR-All}), and on the top $37\%$ images~(\emph{PSNR-Top}).} \label{table:fl_rounds}
	
	\newcommand{\threecol}[1]{\multicolumn{2}{c}{#1}}
	\newcommand{\fivecol}[1]{\multicolumn{5}{c}{#1}}
	\newcommand{\ninecol}[1]{\multicolumn{9}{c}{#1}}
	
	\newcommand{\bsz}{Batch Size~}
	\newcommand{\certified}{{CR(\%)}}
	
	\renewcommand{\arraystretch}{1.2}
	
	\newcommand{\ccellt}[2]{\colorbox{#1}{\makebox(20,8){{#2}}}}
	\newcommand{\ccellc}[2]{\colorbox{#1}{\makebox(8,8){{#2}}}}
	\newcommand{\ccells}[2]{\colorbox{#1}{\makebox(55,8){{#2}}}}
	
	\newcommand{\temp}[1]{\textcolor{red}{#1}}
	\newcommand{\noopcite}[1]{} 
	
	\newcommand{\skiplen}{0.004\linewidth} 
	\newcommand{\rlen}{0.01\linewidth} 
	
	\resizebox{0.6\linewidth}{!}{
	\begingroup
	\setlength{\tabcolsep}{5pt} %
	\begin{tabular}{@{}l crrrr@{}} \toprule
		
		\#Rounds & Learning Rate & Rec (\%) & PSNR-Top $\uparrow$ & PSNR-All $\uparrow$ \\ 
		\midrule
		0 & \num{1e-4} & $94.2$ & $\bm{31.9\pm 1.7}$ & $\bm{28.2\pm 4.3}$ \\
		1 & \num{1e-4} & $\bm{94.6}$ & $31.8\pm 1.7$ & $\bm{28.2\pm 4.3}$ \\
		2 & \num{1e-4} & $94.4$ & $31.1\pm 1.6$ & $27.3\pm 4.2$ \\
		3 & \num{1e-4} & $89.2$ & $27.2\pm 1.7$ & $23.6\pm 3.6$ \\	
		4 & \num{1e-4} & $25.8$ & $20.1\pm 1.5$ & $17.7\pm 2.3$ \\	
		5 & \num{1e-4} & $6.1$  & $17.5\pm 1.5$ & $14.9\pm 2.3$ \\	
		1 & \num{5e-4} & $91.4$ & $29.0\pm 1.7$ & $25.1\pm 4.0$ \\
		\bottomrule
	\end{tabular}
	\endgroup
}
\end{table}

\begin{figure}[!ht]
	\centering
	\includegraphics[width=0.95\linewidth]{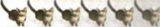}
	\caption{Example reconstructions of \tool{} after training the client model for different number of FL rounds using 8 clients without secure aggregation with learning rate \num{1e-4}. Left to right: Ground truth, 1 round, 2 rounds, 3 rounds, 4 rounds, 5 rounds.}
	\label{fig:fl_rounds}
\end{figure}  

Throughout this paper, we assumed that our attack is mounted at the beginning of the training procedure. 
In this section instead, we show the results of using \tool{}'s decoder to reconstruct images from gradients computed on models that were trained for several federated learning rounds from the malicious state chosen by \tool{}. We show the results quantitatively in \cref{table:fl_rounds} and qualitatively in \cref{fig:fl_rounds} for different number of rounds (\emph{\#Rounds}) and learning rates \emph{Learning Rate} when training with 8 clients without using secure aggregation.

We observe that our decoder works even after several rounds of training of the client model. Further, we observe that our decoder is more robust to a single large step update (One round with learning rate \num{5e-4}) than many smaller updates applied sequentially (Five rounds with learning rate\num{1e-4}) and that each additional communication round results in a small additional drop in quality of the reconstruction. We believe further improvements over these results are achievable if one finetunes our decoder model at each FL round to match the changes of the client encoder, but we leave this as future work.
\subsection{Experiments with Auxiliary Datasets of Different Sizes}\label{sec:app_size_data}
\begin{table}[t]\centering	
	\caption{ Reconstructions on models trained with the Red property and $B=128$ on different percentage $p$ of data in the CIFAR10 trainset. We report the percentage of well-reconstructed images~(\emph{Rec}), the average PSNR and its standard deviation across all reconstructions~(\emph{PSNR-All}), and on the top $37\%$ images~(\emph{PSNR-Top}).} \label{table:data_perc}
	
	\newcommand{\threecol}[1]{\multicolumn{2}{c}{#1}}
	\newcommand{\fivecol}[1]{\multicolumn{5}{c}{#1}}
	\newcommand{\ninecol}[1]{\multicolumn{9}{c}{#1}}
	
	\newcommand{\bsz}{Batch Size~}
	\newcommand{\certified}{{CR(\%)}}
	
	\renewcommand{\arraystretch}{1.2}
	
	\newcommand{\ccellt}[2]{\colorbox{#1}{\makebox(20,8){{#2}}}}
	\newcommand{\ccellc}[2]{\colorbox{#1}{\makebox(8,8){{#2}}}}
	\newcommand{\ccells}[2]{\colorbox{#1}{\makebox(55,8){{#2}}}}
	
	\newcommand{\temp}[1]{\textcolor{red}{#1}}
	\newcommand{\noopcite}[1]{} 
	
	\newcommand{\skiplen}{0.004\linewidth} 
	\newcommand{\rlen}{0.01\linewidth} 
	
	\resizebox{0.4\linewidth}{!}{
		\begingroup
		\setlength{\tabcolsep}{5pt} %
		\begin{tabular}{@{}lrrrrr@{}} \toprule
			
			$p$ & Rec (\%) & PSNR-Top $\uparrow$ & PSNR-All $\uparrow$ \\ 
			\midrule
			5 & $80.6$ & $24.5\pm 1.6$ & $21.5\pm 3.1$ \\
			10 & $92.2$ & $27.9\pm 1.1$ & $24.7\pm 3.4$ \\
			20 & $86.9$ & $30.2\pm 0.9$ & $26.4\pm 4.9$ \\
			33 & $87.5$ & $\bm{31.7\pm 1.1}$ & $27.2\pm 5.5$ \\
			50 & $\bm{95.8}$ & $31.5\pm 1.1$ & $\bm{28.3\pm 3.8}$ \\	
			100 & $93.5$ & $31.1\pm1.2$& $27.8\pm4.1$ \\
			\bottomrule
		\end{tabular}
		\endgroup
	}
\end{table}
\begin{figure}[!ht]
	\centering
	\includegraphics[width=0.9\linewidth]{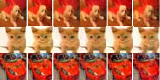}
	\caption{	Example reconstructions of \tool{} trained on different percentage $p$ of CIFAR10 train set. Left to right: Ground truth, 100\%, 50\%, 33\%, 20\%, 10\%, and 5\%.}
	\label{fig:percent}
\end{figure}  

In this section we investigate how the size of the auxiliary dataset used for training \tool{} affects its results. In particular, in \cref{table:data_perc}, we show the results when we used only $p$ percent of the data points in CIFAR10's train set as auxiliary data. As expected of any algorithm based on training, our results generally improve with the number of datapoints available to the attacker. Despite this, we see that our method is very sample efficient, as we successfully reconstruct data from $>80\%$ of client batches even when training on mere 2500 training data samples ($p=5\%$). We further show qualitative comparison between the models in \cref{fig:percent}, where we confirm the quality of reconstructions degrades when less data is available, especially for very small $p$, yet for all $p$ the images remain recognizable regardless.
\subsection{Results Under Client Heterogeneity}\label{app:hetero}
\begin{table}[t]\centering	
	\caption{ Reconstructions for clients with different data heterogeneity levels $\alpha$ on a CIFAR10 model trained with the Red property on $B=128$. We report the percentage of well-reconstructed images~(\emph{Rec}), the average PSNR and its standard deviation across all reconstructions~(\emph{PSNR-All}), and on the top $37\%$ images~(\emph{PSNR-Top}).} \label{table:hetero}
	
	\newcommand{\threecol}[1]{\multicolumn{2}{c}{#1}}
	\newcommand{\fivecol}[1]{\multicolumn{5}{c}{#1}}
	\newcommand{\ninecol}[1]{\multicolumn{9}{c}{#1}}
	
	\newcommand{\bsz}{Batch Size~}
	\newcommand{\certified}{{CR(\%)}}
	 
	\renewcommand{\arraystretch}{1.2}
	
	\newcommand{\ccellt}[2]{\colorbox{#1}{\makebox(20,8){{#2}}}}
	\newcommand{\ccellc}[2]{\colorbox{#1}{\makebox(8,8){{#2}}}}
	\newcommand{\ccells}[2]{\colorbox{#1}{\makebox(55,8){{#2}}}}
	
	\newcommand{\temp}[1]{\textcolor{red}{#1}}
	\newcommand{\noopcite}[1]{} 
	
	\newcommand{\skiplen}{0.004\linewidth} 
	\newcommand{\rlen}{0.01\linewidth} 
	
	\resizebox{0.4\linewidth}{!}{
		\begingroup
		\setlength{\tabcolsep}{5pt} %
		\begin{tabular}{@{}lrrrrr@{}} \toprule
			
			$\alpha$ & Rec (\%) & PSNR-Top $\uparrow$ & PSNR-All $\uparrow$ \\ 
			\midrule
			0.1 & $93.9$ & $28.1\pm 1.0$ & $25.3\pm 3.3$ \\
			0.2 & $94.4$ & $28.7\pm 1.1$ & $26.0\pm 3.3$ \\
			0.3 & $94.3$ & $29.1\pm 1.0$ & $26.2\pm 3.4$ \\
			0.4 & $95.1$ & $29.2\pm 1.1$ & $26.4\pm 3.4$ \\
			0.5 & $95.0$ & $29.4\pm 1.0$ & $26.6\pm 3.4$ \\
			0.6 & $\bm{96.2}$ & $29.5\pm 1.0$ & $26.7\pm 3.2$ \\
			0.7 & $95.8$ & $29.4\pm 1.1$ & $26.6\pm 3.3$ \\
			0.8 & $95.8$ & $29.7\pm 1.1$ & $\bm{27.0\pm 3.3}$ \\
			0.9 & $95.8$ & $29.6\pm 1.0$ & $26.9\pm 3.3$ \\
			1.0 & $96.1$ & $\bm{29.8\pm 1.1}$ & $\bm{27.0\pm 3.3}$ \\
			\bottomrule
		\end{tabular}
		\endgroup
	}
\end{table}
Next, we look at how our method is affected by the level of client data heterogeneity. In particular, we take our CIFAR10 model trained with $B=128$ and the Red property, and we evaluate its performance on clients with different level of non-IID data. To simulate non-IID data, we sample from a Dirichlet distribution with parameter $\alpha$, to determine each client's label distribution and randomly sample the client data according to this. In this setting, $\alpha$ close to $0$ means that each client only holds data from few classes. We show the results in \cref{table:hetero}, where we show that while heterogeneity slightly degrades our performance, \tool{} is very robust to heterogeneity, as even severe heterogeneity levels like $\alpha=0.1$ produce above $90\%$ attack success rate with average PSNR $>25$.
\subsection{Results on the ResImageNet Dataset}\label{sec:app_imagenet_full}
\begin{figure}[!ht]
	\centering
	\includegraphics[width=0.7\linewidth]{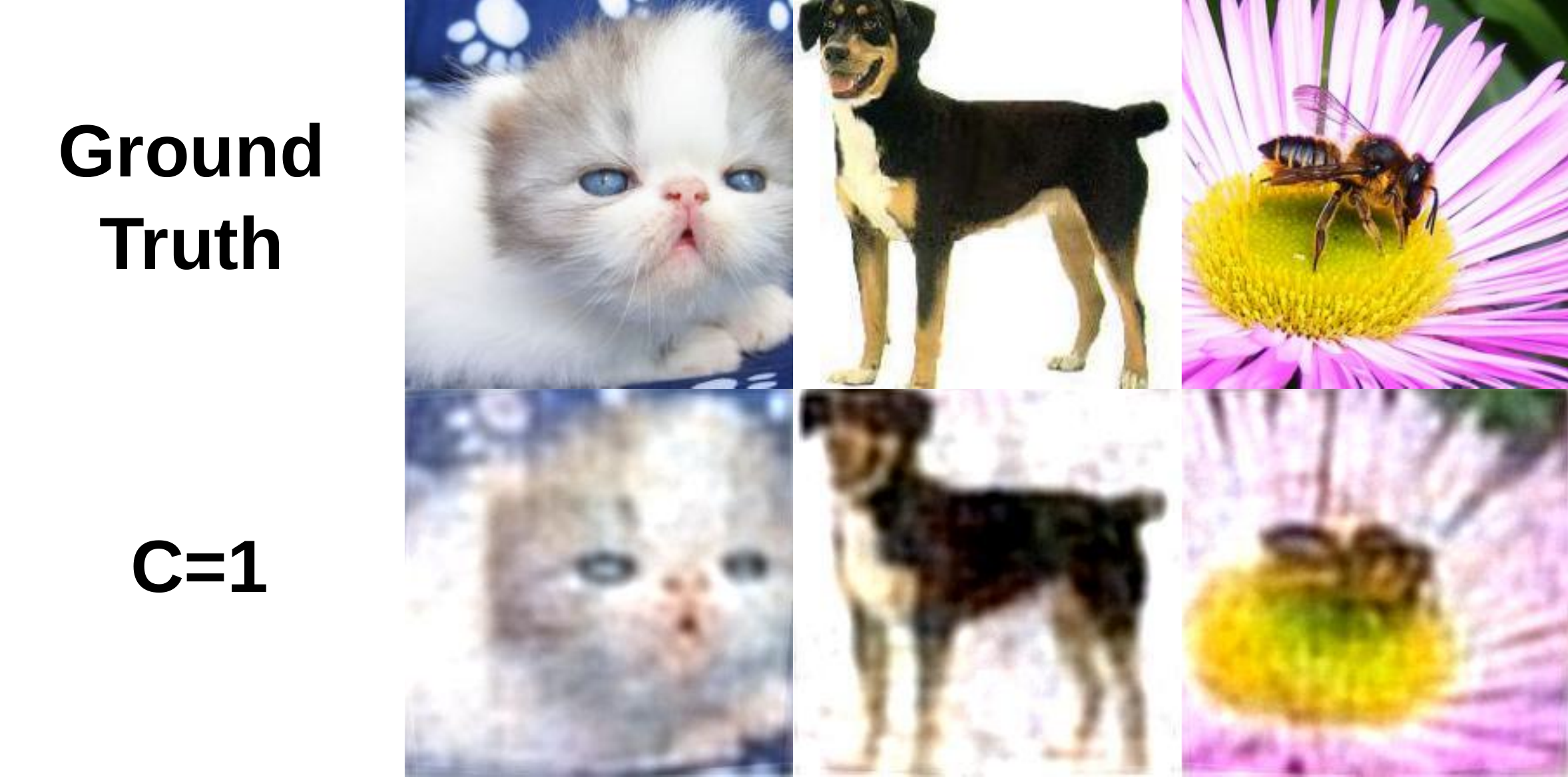}
	\caption{Example reconstructions of \tool{} on ResImageNet for $B=64$ and the Bright property.}
	\label{fig:imagenet_orig}
\end{figure}  

In this section, we show quantitative and qualitative results from applying \tool{} on the ResImageNet dataset (Restricted ImageNet \citep{resimagenet}, a subset of ImageNet with $9$ superclasses). Our setup is similar to the experiments presented for ImageNet in the main paper, \ie we use batch size $B=64$, the Bright property, and our U-Net decoder architecture (\cref{sec:app_unet}). Further, we also use the same hyperparameters (\cref{sec:app_hyperparams}), except for two small changes: (i) we execute the pretraining stage on the images in CIFAR10, instead of the downsized version of the images in ImageNet, avoiding pretraining on the full $1000$ classes, (ii) we train for 370 epochs instead, with 400 gradient descent steps per epoch. 

Under these settings, \tool{} is able to recover $77\%$ of images with average PSNR of $20.6\pm3.7$ and PSNR Top of $23.8\pm1.4$. Examples are shown in \cref{fig:imagenet_orig}, where we see that images are clearly recognizable and accurate in terms of object positions.
\subsection{Results Under Data Domain Shifts}\label{app:isic2019}
\begin{figure}[!ht]
	\centering
	\includegraphics[width=0.77\linewidth]{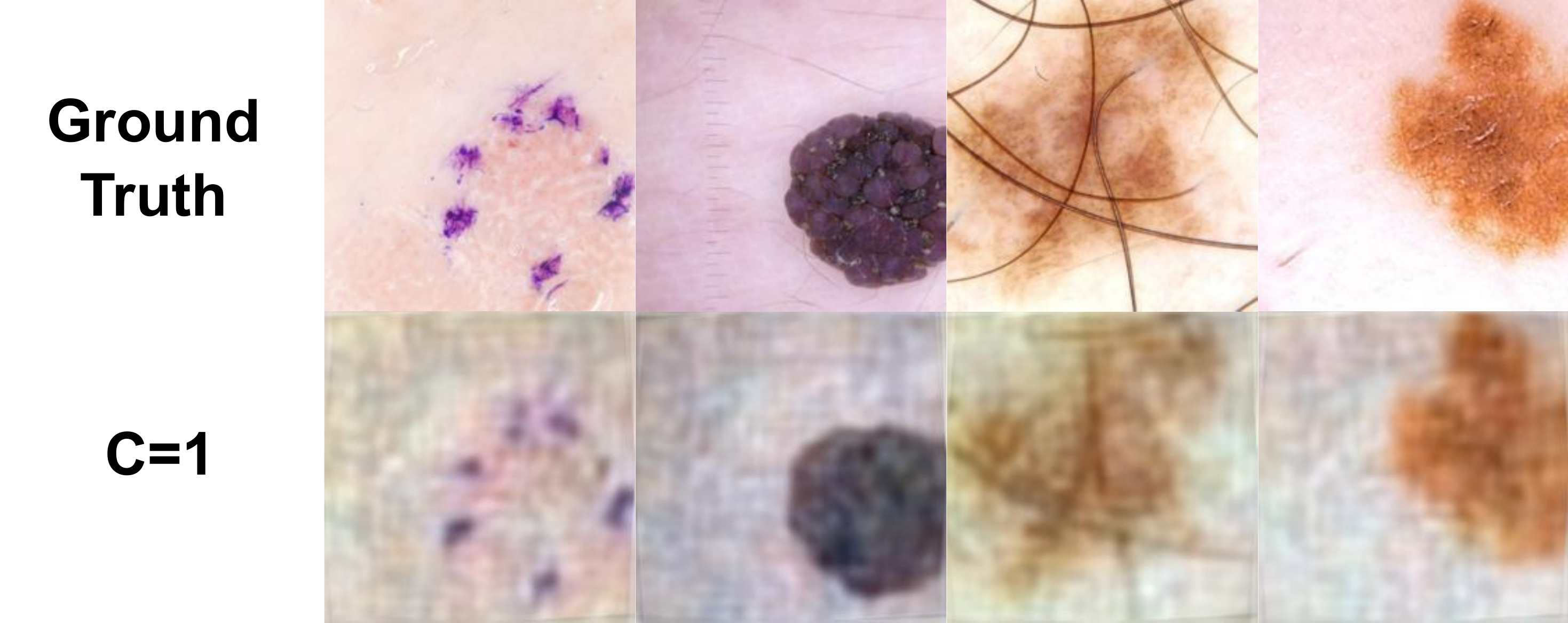}
	\caption{Example reconstructions of \tool{} trained on ImageNet with $B=64$ and the Bright property and applied on ISIC2019.}
	\label{fig:moles}
\end{figure}  

In this section, we first describe some implementation details for our ISIC2019 domain shift experiment originally presented in \cref{table:shifts} that were omitted from the main text for brevity. In particular, to apply our CIFAR10 network trained with the Red property on $B=128$ to the ISIC2019 dataset, we resized the ISIC2019 images so that their larger side is 350 pixels, then applied a random $224\times224$ crop, followed by another downsizing of the image to $32\times32$. Since the domain naturally includes a lot of red images, as the data represents pictures of skin conditions, we saw that our method generated too brightly red images. To avoid this issue, we applied the fix presented in \cref{app:corruptions} for the \textit{Fog} and \textit{Contrast} corruptions. Note that as before, the fix requires no additional knowledge about the color distributions of the images in the ISIC2019 dataset.

Next, we explore a high-resolution version of our domain shift experiment presented in \cref{table:shifts}. In particular, we use the ImagetNet network trained in \cref{sec:app_imagenet_full} and apply it to the ISIC2019 dataset prepared like above but with final resolution $224\times224$ and only $B=64$. We successfully recover data from $76.3\%$ of client batches with average PSNR $20.0\pm 1.8$ and PSNR Top of $21.7\pm 0.8$. We further show example reconstructions of the images in \cref{fig:moles}. Similarly to \cref{table:shifts}, we obtain slightly worse reconstructions compared to the dataset used for the training, but we see that the skin conditions in \cref{fig:moles} are clearly identifiable from our reconstructed images. 
\section{U-Net-based Image Reconstructor}\label{sec:app_unet}
\begin{figure}[t]
	\centering
	\includegraphics[width=0.9 \textwidth]{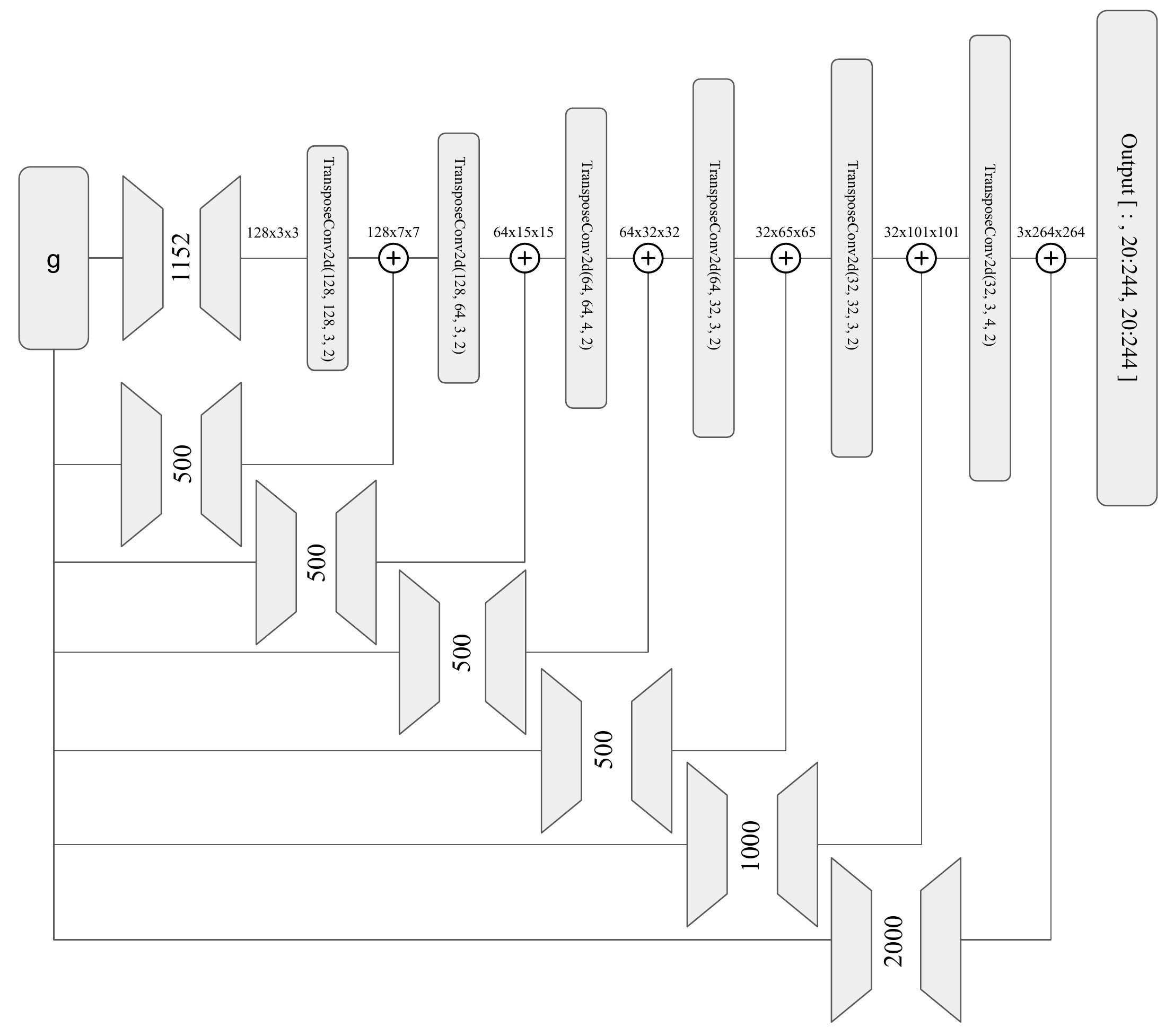}
	\caption{The architecture of our U-Net-based reconstructor $r$ used in our ImageNet experiments. Here $\bm{g}$ is the randomly subsampled model gradient and the output is the resulting reconstructed image.}
	\label{fig:unet}
	\vspace{-1.5em}
\end{figure}
We explain the architecture of our image reconstructor $r$ used in our ImageNet experiments in \cref{sec:experiments}. Our architecture is inspired by the decoder portion of a U-Net~\cite{unet}, which has been demonstrated to be a memory-efficient architecture for generating images.

We show our architecture in \cref{fig:unet}. In the figure, $\bm{g}$ depicts our model's gradient subsampled randomly so that only $3\%$ of its entries are kept (See \cref{sec:app_grad_sample}). There are two main differences between our $r$  in \cref{fig:unet} and the original 6-layer U-Net architecture. First, we use no activation functions, thus creating a (sparse) linear reconstructor function $r$. This allows us, similarly to our CIFAR10/100 experiments, to combine $r$ and $d$ into a single linear layer, whose bias becomes the target for our filtered-out inputs $\bm{X}_\text{nul}$. Second, as our architecture does not include U-Net-style encoder, the U-connections of our reconstructor are substituted by pairs of linear layers with a bottleneck in the middle applied on $\bm{g}$. In \cref{fig:unet}, we depict the bottleneck sizes and transposed convolution sizes, as well as the intermediate output sizes of the different layers. The bottlenecks ensure that the memory efficiency of our method is preserved and are inspired by the intuition that the U-connections only need to provide high-frequency content which can live in a much lower-dimensional subspace. Finally, we note that for the purpose of pretraining, we used the first 3 channels of the third transposed convolution layer as our downsized image output and that our transposed convolution stack produces images of size $264\times264$, which we then center-crop to produce our ImageNet-sized final output.
\section{Hyperparameters} \label{sec:app_hyperparams}
In this section, we provide more details about the exact hyperparameters used in our experiments in \cref{sec:experiments}.
We implemented \tool{} in Pytorch 1.13. Throughout our experiments, we used the Adam optimizer with a learning rate of $0.0001$.
For CIFAR10/100, we trained between 500 and 1000 epochs, where an epoch is defined to be 1000 sampled batches from our trainset. To stabilize our training convergence, we adopted gradient accumulation and, thus, updated our modules' parameters only once every $10$ gradient steps amounting to 100 gradient descent steps per epoch for those datasets. For ImageNet, we additionally execute a pretraining stage on a downsized version of the training images allowing the network to first learn to recover large details in the image before recovering finer details. For ImageNet, we train with the original $\ell_2$-based loss $\mathcal{L}_\text{rec}$ for the first 200 epochs, followed by 300 epochs of using $\ell_1$ version of it, resulting in better visual quality of the reconstruction. At each epoch we only use $4000$ randomly sampled batches instead of the full dataset to reduce computational complexity. 

For faster convergence and better balance in the optimized objective $\mathcal{L} = \mathcal{L}_{\text{rec}} + \alpha\cdot\mathcal{L}_{\text{nul}}$, we adopted a schedule for the hyperparameter $\alpha$, following an exponential curve of the epoch $\kappa$.\\ The schedule is defined as: $\alpha(\kappa) = min(|B|,2^{\beta(\kappa)})$, where $\beta(\kappa) = \frac{(K - \kappa)\beta_{0} + \kappa \beta_{1}}{K}$ linearly interpolates between $\beta_0$ and $\beta_1$ across the total number of epochs $K$ with $(\beta_0,\beta_1)$ set to $(-2,log_2{|B|})$. For ImageNet, we set $(\beta_0,\beta_1)$ to $(-5,5.3)$ to allow for better reconstruction earlier in the training process.
Finally, we point out that in the multi-client setting, we estimate the cumulative density functions before training for the first and second-highest brightness in a batch on $20000$ randomly sampled batches from the trainset (see \cref{sec:app_secagg}).

\section{Implementation Details} \label{sec:app_imp_detail}

\subsection{Subsampled gradients}\label{sec:app_grad_sample}
In order to save memory and computation, we use only part of the entries in our full model gradient $\bm{g}$ to construct our intermediate disaggregation space $\mathbb{R}^{n_d}$. In particular, we randomly sample $0.1\%$ of the gradient entries of each of the model's parameters while ensuring that at least $8400$ entries per parameter are sampled for our CIFAR10/100 experiments, and $2\%$ and at least $9800$ entries for our ResImageNet experiments. This results in $1.6\%$ of the total gradient entries for CIFAR10/100 and $3.0\%$ for ResImageNet. We theorize that we are able to reconstruct nearly perfectly with such a small percent of the gradient entries because there is large redundancy in the information different gradient entries provide. 

\subsection{Efficiently Computing the Disaggregation Loss $\mathcal{L}_\text{null}$}
Computing $\mathcal{L}_\text{null}$ directly for large batch sizes $B$ takes a lot of memory due to the need to store $\bm{g}_i$ for all $i$ in the large set $I_{\text{nul}}$. Note that the reason for this is that we want to enforce all of the individual gradient $\bm{g}_i$ to fall in the null space of $\theta_d$ separately. In practice, to save space, we enforce the same condition by computing the surrogate:
\begin{equation*}
	\widehat{\mathcal{L}}_\text{nul}=\|\,\frac{1}{|I_{\text{nul}}|}\sum_{i \in I_{\text{nul}}}d(\vg_i)\,\|^2_2 + \|\,d(\vg_{j\sim I_{\text{nul}}})\,\|^2_2,
\end{equation*}
where the first part of the equation enforces the mean gradient, and the second part enforces a different randomly chosen gradient at every SGD  to both approach $0$. This, in practice, has a similar result to the original loss $\mathcal{L}_\text{null}$ in that, over time, all gradients in $I_{\text{nul}}$ go to $0$. 

\begin{table*}[t]\centering
	
	\caption{Large batch reconstruction on the bright and red properties from batches of different sizes $B$ on CIFAR10. We report several additional quality of reconstruction metrics---the average MSE and its standard deviation across all reconstructions~(\emph{MSE All}), and on the top $37\%$ images~(\emph{MSE Top}), as well as, the average LPIPS and its standard deviation across all reconstructions~(\emph{LPIPS All}), and on the top $37\%$ images~(\emph{LPIPS Top}).} \label{table:single_c10_full_add}
	
	\newcommand{\threecol}[1]{\multicolumn{4}{c}{#1}}
	\newcommand{\fivecol}[1]{\multicolumn{5}{c}{#1}}
	\newcommand{\ninecol}[1]{\multicolumn{9}{c}{#1}}
	
	\newcommand{\bsz}{Batch Size~}
	\newcommand{\certified}{{CR(\%)}}
	
	\renewcommand{\arraystretch}{1.2}
	
	\newcommand{\ccellt}[2]{\colorbox{#1}{\makebox(20,8){{#2}}}}
	\newcommand{\ccellc}[2]{\colorbox{#1}{\makebox(8,8){{#2}}}}
	\newcommand{\ccells}[2]{\colorbox{#1}{\makebox(55,8){{#2}}}}
	
	\newcommand{\temp}[1]{\textcolor{red}{#1}}
	\newcommand{\noopcite}[1]{} 
	
	\newcommand{\skiplen}{-0.4em} 
	\newcommand{\rlen}{0.01\linewidth} 
	
	\resizebox{\linewidth}{!}{
		\begingroup
		\setlength{\tabcolsep}{5pt}
		\begin{tabular}{@{}l rrrr p{\skiplen}  rrrr@{}} \toprule
			
			& \threecol{CIFAR10, Red} && \threecol{CIFAR10, Bright}\\
			
			\cmidrule(l{5pt}r{5pt}){2-5} \cmidrule(l{5pt}r{5pt}){7-10}
			
			$B$&  MSE Top $\downarrow$ & MSE All $\downarrow$ & LPIPS Top $\downarrow$ & LPIPS All $\downarrow$ &&   MSE Top $\downarrow$ & MSE All $\downarrow$ & LPIPS Top $\downarrow$ & LPIPS All $\downarrow$ \\ \midrule
			64 & $0.0009\pm0.0002$ & $0.0068\pm0.0165$ & \bm{$0.039\pm0.012$}	& $0.128\pm0.194$ && $0.0007\pm0.0002$ & $0.0048\pm0.0099$ & \bm{$0.046\pm0.026$} & \bm{$0.104\pm0.099$}\\
			128 & \bm{$0.0008\pm0.0002$} & $0.0032\pm0.0072$ & $0.050\pm0.014$	& $0.096\pm0.089$ && $0.0007\pm0.0002$ & \bm{$0.0031\pm0.0060$} & $0.057\pm0.025$ & \bm{$0.104\pm0.081$}\\
			256 & \bm{$0.0008\pm0.0002$} & \bm{$0.0029\pm0.0053$} & $0.052\pm0.017$	& \bm{$0.095\pm0.078$} && \bm{$0.0006\pm0.0002$} & $0.0033\pm0.0075$ & $0.054\pm0.025$ & $0.109\pm0.087$ \\
			512 & $0.0010\pm0.0002$ & $0.0035\pm0.0051$ & $0.089\pm0.025$	& $0.141\pm0.081$ && $0.0024\pm0.0007$ & $0.0070\pm0.0091$ & $0.168\pm0.058$ & $0.204\pm0.089$\\
			\bottomrule
		\end{tabular}
		\endgroup
	} 
\end{table*}
\subsection{Trainset data augmentation}
For the purpose of training our encoder-decoder framework, we observed data augmentation of our auxiliary dataset is crucial, especially for large batch sizes $B$. We theorize that the reason for this is the lack of diversity in the reconstruction samples $\bm{X}_\text{rec}$. In particular, as $B$ grows, an increasingly smaller set of images are selected to be the brightest or darkest of any batch sampled from the training set. To this end, when sampling our training batches for CIFAR10/100, we first apply random ColorJitter with brightness, contrast, saturation, and hue parameters $0.2$, $0.1$, $0.1$, and $0.05$, respectively, followed by random horizontal and vertical flips, and random rotation at $N*90 + \epsilon$ degrees, where $N$ is a random integer and $\epsilon$ is chosen uniformly at random on $[-5,5]$. For ResImageNet,
we additionally do a random cropping of the original image to the desired size of $224\times224$ before the other augmentations.

\subsection{Trainset batch augmentation}
 As detailed in \cref{sec:app_secagg}, when mounting \tool{} in the secure aggregation setting we select $\tau$ based on a mix of the global and local distributions of $m$. As noted in \cref{sec:attack:components}, the probability of attack success with an optimal threshold $\tau$ based on the global distribution of $m$ is $\frac{1}{e}$ in the limit of the number of images being aggregated. We expect for large $B$, therefore, \tool{} to also successfully reconstruct only for $\approx\frac{1}{e}$ of the securely-aggregated batches, forcing us to avoid training on the rest $1-\frac{1}{e} > \frac{1}{2}$ of the securely-aggregated batches, which act as a strong noise during training and prevent convergence. This, in turn, results in rejecting training on a big portion of our sampled client batches.

To address this sample inefficiency, we use batch augmentation during training to transform the client batches to ones with a desired brightness distribution. The batch augmentation simply consists of adjusting the brightnesses of individual images within each batch. We do two types of batch augmentations based on two different distributions---one where it contains precisely one image in the batch with brightness above the threshold $\tau$ and another where precisely zero images in the batch have brightness above the threshold $\tau$. We alternate the two augmentations at each step of the training procedure.
To achieve the distributions, we adjust all image brightness within a batch using a heuristic method. The method first adjusts the brightness of the most bright image (the least bright image in the case of the dark image property) such that it lands on the desired side of the threshold $\tau$. However, as after adjusting the image, the brightnesses within the batch are no longer normalized, we then need to renormalize the batch, resulting in a new batch brightness distribution. If our new distribution is as desired, we stop. Otherwise, we iterate the process until convergence.
 \subsection{Details on Prior Work Comparison}\label{sec:app_fishing_comp_details} 
 In this section, we provide further details about the exact setting in which we do the comparison against prior work in \cref{table:fishing_comp} and \cref{fig:dsnr} in the main text. We focus on the feature fishing variant of \citet{fishing}, that targets batches containing a large percentage of repeated labels, as this setting was shown in prior work~\citep{nvidia,aaai} to be significantly harder to solve.
For the comparison itself, we evaluate both the Fishing and \tool{} models on client batches of the same class. This setting favors \citet{fishing} over \tool{}, as the Fishing's weights are specifically adapted to it and the \tool{} model was only trained on batches with randomly selected mixed labels, as in the rest of the paper. We select D-SNR threshold of 5 for the experiment in \cref{table:fishing_comp} based on \cref{fig:dsnr}.
For the comparison against Zhang23~\citep{zhang23} and LOKI~\citep{mandrake}, we observe they have practically infinite T-SNR, making them trivially detectable.
\section{Additional Measurements of the Quality of Our Reconstructed Images}\label{sec:app_ssim}
\begin{table*}[t]\centering
	
	\caption{Large batch reconstruction on the bright and red properties from batches of different sizes $B$ on CIFAR100. We report several additional reconstruction quality metrics---the average MSE and its standard deviation across all reconstructions~(\emph{MSE All}), and on the top $37\%$ images~(\emph{MSE Top}), as well as, the average LPIPS and its standard deviation across all reconstructions~(\emph{LPIPS All}), and on the top $37\%$ images~(\emph{LPIPS Top}).} \label{table:single_c100_full_add}
	
	\newcommand{\threecol}[1]{\multicolumn{4}{c}{#1}}
	\newcommand{\fivecol}[1]{\multicolumn{5}{c}{#1}}
	\newcommand{\ninecol}[1]{\multicolumn{9}{c}{#1}}
	
	\newcommand{\bsz}{Batch Size~}
	\newcommand{\certified}{{CR(\%)}}
	
	\renewcommand{\arraystretch}{1.2}
	
	\newcommand{\ccellt}[2]{\colorbox{#1}{\makebox(20,8){{#2}}}}
	\newcommand{\ccellc}[2]{\colorbox{#1}{\makebox(8,8){{#2}}}}
	\newcommand{\ccells}[2]{\colorbox{#1}{\makebox(55,8){{#2}}}}
	
	\newcommand{\temp}[1]{\textcolor{red}{#1}}
	\newcommand{\noopcite}[1]{} 
	
	\newcommand{\skiplen}{-0.4em} 
	\newcommand{\rlen}{0.01\linewidth} 
	
	\resizebox{\linewidth}{!}{
		\begingroup
		\setlength{\tabcolsep}{5pt}
		\begin{tabular}{@{}l rrrr p{\skiplen}  rrrr@{}} \toprule
			
			& \threecol{CIFAR100, Red} && \threecol{CIFAR100, Bright}\\
			
			\cmidrule(l{5pt}r{5pt}){2-5} \cmidrule(l{5pt}r{5pt}){7-10}
			
			$B$&  MSE Top $\downarrow$ & MSE All $\downarrow$ & LPIPS Top $\downarrow$ & LPIPS All $\downarrow$ &&   MSE Top $\downarrow$ & MSE All $\downarrow$ & LPIPS Top $\downarrow$ & LPIPS All $\downarrow$ \\ \midrule
			64 & $0.0007\pm0.0002$ & $0.0021\pm0.0046$ & $0.034\pm0.010$ & $0.056\pm0.059$ && $0.0006\pm0.0002$ & $0.0028\pm0.0047$ & $0.050\pm0.023$ & $0.094\pm0.075$\\
			128  & $0.0007\pm0.0001$ & $0.0019\pm0.0039$ & \bm{$0.031\pm0.011$} & \bm{$0.051\pm0.061$} && $0.0010\pm0.0003$ & $0.0035\pm0.0045$ & $0.112\pm0.035$ & $0.164\pm0.076$\\
			256 & $0.0008\pm0.0002$ & $0.0022\pm0.0047$ & $0.047\pm0.014$ & $0.079\pm0.064$ && \bm{$0.0003\pm0.0001$} & \bm{$0.0017\pm0.0031$} & \bm{$0.042\pm0.019$} & \bm{$0.091\pm0.068$}\\
			512 &\bm{$0.0005\pm0.0001$} & \bm{$0.0014\pm0.0025$} & $0.034\pm0.009$ & $0.057\pm0.049$ && $0.0006\pm0.0002$ & $0.0032\pm0.0043$ & $0.077\pm0.032$ & $0.148\pm0.085$\\
			\bottomrule
		\end{tabular}
		\endgroup
	} 
\end{table*}

\begin{table*}[t]\centering

	\caption{Reconstruction from securely aggregated updates on the bright and dark properties using different numbers of clients $C$ on CIFAR10, for different total numbers of images. We report additional reconstruction quality measures---the average MSE (\emph{MSE Top}) and LPIPS (\emph{LPIPS Top}) and their respective standard deviations on the top $37\%$ images.} \label{table:secure_add}
	
	\newcommand{\threecol}[1]{\multicolumn{2}{c}{#1}}
	\newcommand{\fivecol}[1]{\multicolumn{5}{c}{#1}}
	\newcommand{\ninecol}[1]{\multicolumn{9}{c}{#1}}
	
	\newcommand{\bsz}{Batch Size~}
	\newcommand{\certified}{{CR(\%)}}
	
	\renewcommand{\arraystretch}{1.2}
	
	\newcommand{\ccellt}[2]{\colorbox{#1}{\makebox(20,8){{#2}}}}
	\newcommand{\ccellc}[2]{\colorbox{#1}{\makebox(8,8){{#2}}}}
	\newcommand{\ccells}[2]{\colorbox{#1}{\makebox(55,8){{#2}}}}
	
	\newcommand{\temp}[1]{\textcolor{red}{#1}}
	\newcommand{\noopcite}[1]{} 
	
	\newcommand{\skiplen}{-0.1em} 
	\newcommand{\rlen}{0.01\linewidth} 
	
	\resizebox{1.0 \linewidth}{!}{
		\begingroup
		\setlength{\tabcolsep}{5pt}
		\begin{tabular}{@{}l rr p{\skiplen}  rr p{\skiplen} rr p{\skiplen}  rr p{\skiplen}  rr  p{\skiplen}  rr@{}} \toprule
			
			& \threecol{$C=4$, Dark} && \threecol{$C=4$, Bright} && \threecol{$C=8$, Dark} && \threecol{$C=8$, Bright}\\
			
			\cmidrule(l{5pt}r{5pt}){2-3} \cmidrule(l{5pt}r{5pt}){5-6} \cmidrule(l{5pt}r{5pt}){8-9} \cmidrule(l{5pt}r{1pt}){11-12}
			
			\#Imgs&  MSE Top $\downarrow$ & LPIPS Top  $\downarrow$ && MSE Top $\downarrow$ & LPIPS Top  $\downarrow$ &&   MSE Top $\downarrow$ & LPIPS Top  $\downarrow$ &&   MSE Top $\downarrow$ & LPIPS Top  $\downarrow$ \\ \midrule
			64 & $0.0020\pm0.0013$ & \bm{$0.110\pm0.052$} && $0.0027\pm0.0022$ & \bm{$0.091\pm0.060$} && $0.0031\pm0.0023$ & \bm{$0.111\pm0.059$} && \bm{$0.0026\pm0.0021$} & $0.123\pm0.073$ \\
			128 & $0.0018\pm0.0013$ & $0.130\pm0.050$ && $0.0028\pm0.0018$ & $0.105\pm0.064$ && $0.0028\pm0.0023$ & $0.118\pm0.070$ && $0.0031\pm0.0026$ & \bm{$0.111\pm0.075$} \\
			256 & \bm{$0.0012\pm0.0008$} & $0.130\pm0.045$ && \bm{$0.0022\pm0.0012$} & $0.113\pm0.049$ && $0.0023\pm0.0014$ & $0.130\pm0.060$ && $0.0035\pm0.0022$ & $0.164\pm0.085$ \\
			512 & \bm{$0.0012\pm0.0008$} & $0.142\pm0.054$ && $0.0030\pm0.0013$ & $0.208\pm0.064$ && \bm{$0.0015\pm0.0010$} & $0.159\pm0.054$ && $0.0029\pm0.0017$ & $0.170\pm0.074$ \\
			\bottomrule
		\end{tabular}
		\endgroup
	}
\end{table*}

In \cref{sec:experiments} and \cref{sec:app_full_exp}, we focussed on reporting the quality of our image reconstructions in terms of the popular PSNR image quality metric. In this section, we provide additional image quality measurements in terms of the mean square error of the individual pixels (\emph{MSE}) and the learned perceptual image patch similarity (\emph{LPIPS}) metrics. We present the additional measurements for all large batch experiments from \cref{sec:app_full_exp} in \cref{table:single_c10_full_add} and \cref{table:single_c100_full_add} for CIFAR10 and CIFAR100, respectively. Further, in \cref{table:secure_add}, we present the additional measurements for the multi-client experiments originally presented in \cref{sec:experiments}.
The additional measurements reinforce our observations from \cref{sec:experiments} that \tool{} consistently reconstructs client data well.

\fi

\end{document}